\documentclass{article}

\usepackage{arxiv}

\usepackage[utf8]{inputenc}
\usepackage[T1]{fontenc}
\usepackage{url}
\usepackage{booktabs}
\usepackage{amsfonts}
\usepackage{nicefrac}
\usepackage{microtype}
\usepackage{graphicx}
\usepackage{amsmath}
\usepackage{subcaption}
\usepackage{enumitem}
\usepackage{fancyhdr}
\usepackage{rotating}
\usepackage{multirow}
\usepackage{esvect}
\usepackage{algorithm}
\usepackage{algpseudocode}
\usepackage{afterpage}
\usepackage{placeins}
\usepackage{bbding}
\usepackage{amsthm}

\usepackage[authoryear]{natbib}   % <-- add option if you want author-year style

\usepackage{hyperref}             % <-- load AFTER natbib
\usepackage{doi}                  % can be after hyperref

\newtheorem{definition}{Definition}

\title{Heuristic Algorithms for the Stochastic Critical Node Detection Problem}

\author{
  Tuguldur Bayarsaikhan\\
  Department of Information and Computer Science\\
  National University of Mongolia, \\
  Ulaanbaatar, Mongolia\\
  \texttt{18B1NUM0267@stud.num.edu.mn}
  \And
  Altannar Chinchuluun\\
  Business School\\
  National University of Mongolia, \\
  Ulaanbaatar, Mongolia\\
  \texttt{altannar@num.edu.mn}
  \And
  Ashwin Arulselvan\\
  Department of Management Science\\
  University of Strathclyde, \\
  Glasgow, United Kingdom\\
  \texttt{ashwin.arulselvan@strath.ac.uk}
  \And
  Panos M. Pardalos\\
  Department of Industrial and Systems Engineering\\
  University of Florida, \\
  Gainesville, FL, USA\\
  \texttt{pardalos@ufl.edu}
}

\date{}

\hypersetup{
pdftitle={A template for the arxiv style},
pdfsubject={q-bio.NC, q-bio.QM},
pdfauthor={David S.~Hippocampus, Elias D.~Striatum},
pdfkeywords={First keyword, Second keyword, More},
}

\begin{document}
\maketitle

\begin{abstract}
In a given graph network, the critical node detection problem finds a subset of nodes whose removal disrupts the network connectivity. Since many real-world systems are naturally modeled as graphs, assessing the vulnerability of the network is essential, with applications in transportation systems, traffic forecasting, epidemic control, and biological networks. In this paper, we consider the stochastic version of the critical node detection problem \cite{Dinh2015Vulnerability}, where the existence of edges is given by certain probabilities. 
% which makes the problem challenging because of the combinatorial explosion. 
We propose heuristics and learning-based methods for the problem and compare them with existing algorithms. Experimental results performed on random graphs from small to larger scales, with edge-survival probabilities drawn from different distributions, demonstrate the effectiveness of the methods. Heuristic methods often illustrate the strongest results with high scalability, while learning-based methods maintain nearly constant inference time as the network size and density grow.
\end{abstract}

% keywords can be removed

\section{Introduction}

Networks are a powerful and flexible representation for describing complex systems. Many real-world problems can be effectively represented as networks, including tracking contagion in epidemic networks \cite{Chaharborj2022COVIDNodes}, improving the reliability of network traffic planning \cite{Wang2024MultimodalTransport}, and modeling protein interactions in biology \cite{Boginski2009PPI}. For example, the network is used for understanding the propagation of ideas and influence in social networks \cite{KKT03}. The research direction has been further extended to study the spread of epidemics within communities and to identify the key individuals whose infections could significantly increase subsequent infections and mitigate disease transmission.

The critical node detection problem (CNDP) was first formalized by \citet{Arulselvan2009SparseCNDP}. The authors proved the problem is $NP$-hard, indicating that as the size of the graph increases, the problem becomes computationally expensive, and they proposed mixed-integer linear programming (MILP) formulations tailored for sparse graphs with a greedy heuristic. However, scaling to larger graphs is difficult because the MILP introduces a significant number of variables and constraints, leading to rapidly growing branch-and-bound trees and excessive memory usage as the counts of nodes and edges increase. To address the challenges, heuristic and metaheuristic methods have been developed that run faster but lack solution quality. Recent surveys, \cite{cappart2023combinatorial,peng2021graph}, highlight the effectiveness of learning-based methods in addressing scalability issues.

Although many studies have investigated the CNDP \cite{Arulselvan2023ICNN}, only a limited number of studies have addressed the evaluation of networks under uncertainty conditions.  In practical applications, uncertainty is inherent because of the dynamic behavior of networks, environmental variability, model or parameter inaccuracy, and data collection or processing errors. Those factors can strongly affect key components of the network, such as critical elements (nodes, edges) of the network, and ignoring them may lead to misleading assessments of network vulnerability. 

The stochastic version of the CNDP over trees is studied by \citet{Hosteins2020StochasticCNP}. The attack on nodes can fail with a certain probability. The authors demonstrate that the problem generalizes to the stochastic critical element detection problem, where edge attacks can fail with a certain probability. Furthermore, the decision version of the problem, in which connection costs are set to one, has been proven to be $NP$-complete. Moreover, \cite{DiSumma2011CNPtrees} shows that the deterministic counterpart of the problem over trees with specified connection cost for each pair of distinct nodes is still $NP$-complete, emphasizing that stochastic uncertainty introduces inherent complexity. In addition, the authors derived linear and nonlinear MILP models and proposed an exact approach, which is built on Benders decomposition that effectively handles a large set of instances.

\citet{Dinh2015Vulnerability} addressed the stochastic critical node detection problem (SCNDP) on probabilistic and real-life graphs. The authors constructed an efficient \emph{Fully Polynomial Time Randomized Approximation Scheme} (FPRAS) to estimate the expected pairwise connectivity (EPC). In their approach, the underlying network is modeled probabilistically, generating numerous graph realizations where each edge exists with a certain probability, contrasting with deterministic graphs, which have fixed topologies. The objective of the model is to minimize the EPC. However, computing EPC exactly is $\#P$-complete due to the exponential number of possible realizations, specifically $2^m$ different scenarios for a graph with $m$ edges. Furthermore, the MILP model they derived also suffers from having exponentially many variables and constraints corresponding to the scenarios. To overcome this challenge, the authors proposed a heuristic method called the rounding the expected graph algorithm (REGA), which solves a reduced MILP with an aggregated set of variables and constraints and utilizes the FPRAS to efficiently estimate EPC without exhaustive enumeration of all scenarios. Similarly, \citet{NengRobust} formulated a variant of the CNDP with nondeterministic connection costs as a robust optimization problem.

\citet{Bayarsaikhan2025} proposed a heuristic method for the SCNDP based on the maximal independent set (MIS). The two problem definitions were provided, including CNDP with uncertain edges and live-edge graph scenarios, borrowed from \cite{KKT03}. The proposed MIS-based heuristic is compared against the established REGA from \cite{Dinh2015Vulnerability} on synthetic random graphs under different probability settings, including uniform and heterogeneous distributions. Empirically, the MIS approach shows competitive results against REGA while being markedly faster, even when integrated with a local search refinement for the solution quality. They conclude that MIS offers a scalable, practical alternative for tasks involving disruption in stochastic networks, particularly as instance sizes grow.

The structure of the paper is as follows. In \hyperref[sec:problem]{Section~\ref*{sec:problem}}, we review the SCNDP and present two alternative definitions as shown in \citet{Bayarsaikhan2025}. \hyperref[sec:methods]{Section~\ref*{sec:methods}} gives descriptions of the proposed methods in detail. The first method is a greedy algorithm that starts with the empty set and adds nodes one by one. The second algorithm is the greedy algorithm by \citet{Bayarsaikhan2025} that starts with a maximal independent set (MIS). In fact, that is an  extension of the heuristic provided by \cite{Arulselvan2009SparseCNDP} to the stochastic setting. Finally, we employ graph representation learning, which consists of GraphSAGE \cite{HamiltonGraphsage} and our custom-designed edge-probability-aware Graph Attention Network (GAT) layers \cite{veličković2018graph}. \hyperref[sec:experiments]{Section~\ref{sec:experiments}} outlines the experimental results of the proposed algorithms on a wider range of datasets that are more diverse and challenging. The datasets include sparse graph topologies, varying numbers of nodes, and heterogeneous edge survival probabilities that are drawn from different probability distributions, allowing for robust testing of each approach.
Finally, \hyperref[sec:conclusion]{Section~\ref*{sec:conclusion}} concludes the paper and discusses potential directions for future research.

\section{Problem Formulation}\label{sec1}
\label{sec:problem}
Given a graph $G(V,E)$ with a set of vertices $V$, a set of edges $E$, and a budget $k$, the CNDP selects a set of $k$ nodes, $S$, to attack or remove so that the pairwise connectivity of the remaining nodes in $V\setminus S$ is minimized. Let $G_{V\setminus S}$ be the node-induced subgraph with nodes $V\setminus S$.
We now provide the concept of live-edge graphs introduced in \cite{KKT03} for the influence maximization problem. Let $\Omega$ be the set of all possible scenarios of the stochastic graph $\mathcal{G}(V,E,\pi)$, where $V$ is the node set, $E$ is the edge set, and $\pi:E\rightarrow [0,1]$ is the probability function that specifies the probability that an edge exists. We obtain a random scenario $w\in \Omega$ by randomly blocking or activating (to be live) each edge in $G$, resulting in a static subgraph of $G^w=(V, E^w)$, called a live-edge graph.  

\begin{definition}[Live-edge graph]
Given a stochastic graph $\mathcal{G}(V,E,\pi)$, we select each edge $(i,j)\in E$ independently with probability $\pi(i,j)$. The selected edge is declared to be "live" and all other edges are declared to be "blocked". Let us denote the set of live edges by $E^w$ and the resulting graph by $G^w(V, E^w)$ or $G^w$, when it is clear, in scenario $w$. $G^w$ is called the live-edge graph. Each edge $(i,j)\in E^w$ in a live-edge graph indicates that there is an edge between $i$ and $j$ with a probability of $1$ in scenario $w$.
\end{definition}

%Therefore, the pairwise connectivity of the stochastic CNDP problem can be calculated based on the live-edge graph scenarios. 
For scenario $w$, let $\sigma^w$ be the pairwise connectivity of the live-edge $G^w$. Then the expected pairwise connectivity (EPC) of the stochastic graph $G$ is defined as:
\begin{equation*}
\sigma
= \sum_{w\in\Omega} p^{w}\,\sigma^{w}\,.
\end{equation*}
Here, the probability of a scenario $w$ is $p^w=\prod\limits_{(i,j)\in E^w} \pi_{ij}\prod\limits_{(i,j)\in E\setminus E^w} (1-\pi_{ij})$, and the number of all possible live-edge graphs or scenarios is $2^m$, where $m=|E|$.

We extend the above definition of EPC  with respect to a subset of nodes $S$. For each scenario $w$, we denote $\sigma^w(S)$ as the pairwise connectivity of $G^w$ after deleting the node set $S$, i.e., $\sigma^w(S)$ is the pairwise connectivity of the node-induced subgraph $G^w[V\backslash S]$. Then the EPC of the stochastic node-induced subgraph $G[V\backslash S]$ can be defined as follows: 
\begin{equation*}
\sigma(S)
= \sum_{w\in\Omega} p^{w}\,\sigma^{w}(S)\,.
\end{equation*}

\begin{definition}[Stochastic CNDP (SCNP) with uncertain edges \cite{Dinh2015Vulnerability}]
Given a stochastic graph $G(V,E,\pi)$ with a set of vertices $V$, a set of edges $E$, probability function $\pi:E\rightarrow(0,1]$ and a budget $k$, the stochastic CNDP with seeks a subset of $k$ nodes, $S\subset V$, to attack or remove so that the expected pairwise connectivity, $\sigma(S)$ is minimized.
\end{definition}
Our problem can be formulated as the following optimization problem.
\begin{eqnarray*}
\min\limits_ {S\subset V} &&  \sum_{w\in\Omega} p^{w}\,\sigma^{w}(S)\\\
\text{s.t}.&& |S|\leq k.
\end{eqnarray*}

One way to model this is by extending the deterministic formulation introduced in \cite{Arulselvan2009SparseCNDP} to a two-stage stochastic program. We introduce the 0-1 variables $s_i$ for each $i\in V$, indicating whether node $i$ is chosen as critical or not, and for each scenario $w\in\Omega$, we introduce 0-1 connectivity variables $u^w_{ij}$ that model whether nodes $i$ and $j$ are connected. 
\begin{eqnarray*}
 (MIP_F): \qquad  \min &&\sum_{w\in \Omega} p^w \sum_{i,j\in V} u^w_{ij}\\
    \textrm{s.t.} && \sum_{i\in V}s_i \le k \\
    && u^w_{ij}+s_i+s_j\geq 1,~\forall w\in \Omega,~\forall (i,j)\in E^w,\\
    && u^w_{ij}+u^w_{jl}-u^w_{li}\leq 1, ~\forall (i,j,l)\in V,\\
    && u^w_{ij}-u^w_{jl}+u^w_{li}\leq 1, ~\forall (i,j,l)\in V,\\
    && -u^w_{ij}+u^w_{jl}+u^w_{li}\leq 1, ~\forall (i,j,l)\in V,\\
   && s_i \in\{0,1\},~ \forall i\in V,\\
   && u^w_{i,j} \in [0,1],~\forall w\in \Omega,~\forall i\in V. 
\end{eqnarray*}

\section{Methodology}
\label{sec:methods}

    In this section, we present heuristic approaches to the SCNDP. Each approach is explained in detail, and descriptions of the algorithms are provided.

\subsection{REGA}

To assess the performance of our proposed algorithms, we implemented the Rounding the Expected Graph Algorithm (REGA) and the Component Sampling Procedure (CSR) for EPC estimation as described in \cite{Dinh2015Vulnerability}.
    
The component sampling procedure (CSP) efficiently estimates the EPC by sampling graph components instead of entire graphs. It has the advantage of having polynomial-time complexity due to an FPRAS, with the running time bounded by a polynomial in $1/\epsilon, log(1/\delta) $, and the input size, and it significantly reduces computational overhead compared to naive Monte Carlo methods. CSP employs an importance sampling strategy, selecting nodes uniformly and performing localized breadth-first searches to estimate EPC values quickly. To detect critical nodes in large graphs efficiently, we parallelized the computation. We provide the description of CSP in Algorithm \ref{alg:1} for calculating $\sigma$ as described in \cite{Dinh2015Vulnerability}.
    
\begin{algorithm}

        \caption{$(\varepsilon,\delta)$ Component Sampling Procedure (CSP) for calculating $\sigma(S)$}
        \label{alg:csp-sigma}
        \textbf{Input: } Stochastic graph $G=(V,E,\pi)$ and accuracy $(\varepsilon,\delta)$ \\
        \textbf{Output: } Estimator $\widehat{\mathcal{E}}$
        \begin{algorithmic}[1]
            \State $P_E \gets \sum_{e \in E} \pi_e$
            \If{$P_E < \frac{\varepsilon}{2} n^{-2}$}
                \State \Return $\widehat{\mathcal{E}} \gets P_E$
            \EndIf
            \State $C_2 \gets 0$
            \For{$i = 1 \ldots N(\varepsilon,\delta)$}
                \State Select node $u \in V$ uniformly at random
                \State Run BFS from $u$; when exploring edge $(v,w)$, keep it w.p.\ $\pi_{vw}$ (else discard)
                \State Let $S_i$ be number of visited nodes (including $u$)
                \State $C_2 \gets C_2 + (S_i - 1)$
            \EndFor
            \State $\widehat{\mathcal{E}} \gets \dfrac{n\, C_2}{2 N(\varepsilon,\delta)}$
            \State \Return $\widehat{\mathcal{E}}$
        \end{algorithmic}
        \label{alg:1}
\end{algorithm}
We take $N(\epsilon,\delta) \ge 4(e-2)\ln{(\frac{1}{\delta})}\frac{1}{\epsilon^2 EPC}$ in the Algorithm \ref{alg:1}. A lower bound on EPC, $\sigma$, of the stochastic graph $G$ is required to calculate $N(\epsilon,\delta)$, in order to ensure that $\widehat{\mathcal{E}}$ is a $(\epsilon,\delta)$-estimator of $\sigma$, i.e., we have
    \[\Pr\left[(1-\epsilon)\sigma \le\widehat{\mathcal{E}}\le(1+\epsilon) \sigma\right] \ge (1-\delta) \]

REGA is a heuristic algorithm designed to efficiently solve SCNDP. 
It first replaces the huge two‑stage mixed integer program $MIP_F$ with a much smaller linear program by averaging the scenario constraints using their probabilities. After solving this relaxed model, it rounds the fractional solution to choose the $k$ nodes to delete. A local search procedure is applied after deleting $k$ nodes, where nodes in the current deletion set are iteratively swapped with remaining nodes if the swap reduces EPC. We now provide the reduced MIP formulation, $MIP_R$, and the description of REGA in Algorithm \ref{alg:2} as described in \cite{Dinh2015Vulnerability}.

\textit{Reduced Mixed Integer Program}, $MIP_R:$

    \begin{eqnarray*}
            \min && \sum_{i<j} (1 - x_{ij})\\
            \textrm{s.t.} && \sum_{i=1}^{n} s_i \le k \\
            && x_{ij} \le s_i + s_j + 1 - \pi_{ij},\ (i,j)\in E,\\
            && x_{ij} + x_{jk} \ge x_{ik},\ (i,j)\in E,\ k=1,\ldots,n,\\
            && s_i \in \{0,1\},\ x_{ij}\in[0,1],\ i,j=1,\ldots,n
    \end{eqnarray*}
    
    \begin{algorithm}
    \caption{Rounding the Expected Graph Algorithm (REGA)}
    \textbf{Input: } Stochastic graph $G=(V,E,\pi)$, budget $k$ \\
    \textbf{Output: } Deletion set $D$
    \begin{algorithmic}[1]
        \State $D \gets \emptyset$
        \For{$t = 1,\ldots,k$}
            \State Solve the LP relaxation of $\text{MIP}_R$ to obtain fractional solution $s$
            \State $u \gets \arg\max_{i \in V \setminus D} s_i$
            \State $D \gets D \cup \{u\}$ \State $s_u\leftarrow 1$ in $\text{MIP}_R$ 
        \EndFor
        \Repeat
            \State $improved \gets \textbf{false}$
            \ForAll{$(u,v) \in D\times D$}
                %\ForAll{$v \in V \setminus D$}
                    \State Estimate $\sigma(D - \{u\} + \{v\})$ 
                    % using CSP
                    \If{$\sigma(D - \{u\} + \{v\}) < \sigma(D)$}
                        \State $D \gets D - \{u\} + \{v\}$
                        \State $improved \gets \textbf{true}$
                    \EndIf
                %\EndFor
            \EndFor
        \Until{$improved = \textbf{false}$}
        \State \Return $D$
    \end{algorithmic}
    \label{alg:2}
    \end{algorithm}
    
\subsection{Proposed methods}

\subsubsection{Greedy heuristic}

    The greedy algorithm from the empty set is a straightforward heuristic that starts from the empty set and iteratively selects nodes to maximize the EPC reduction immediately until the specified node-removal budget is reached. After the initial selection set of nodes, we apply the same local search procedure used by REGA to ensure fair comparisons.

\begin{algorithm}
\caption{Greedy heuristic}
    \textbf{Input: }Stochastic graph $G(V,E, \pi)$, and budget $k$ \\
    \textbf{Output: }$S$ 
    \begin{algorithmic}[1]
        \State $S \leftarrow \emptyset$
        \For {$i=1,\ldots,k$}
            \State $v\leftarrow \arg\max\limits_{j\in V\setminus S} \sigma(S) - \sigma(S\cup j)$
            \State $S\leftarrow S\cup v$ 
        \EndFor
    \end{algorithmic}
\end{algorithm}
In this algorithm, to estimate $\sigma(S)$, we use Algorithm 1.

\begin{figure}[h]
    \centering
    \includegraphics[width=0.6\textwidth]{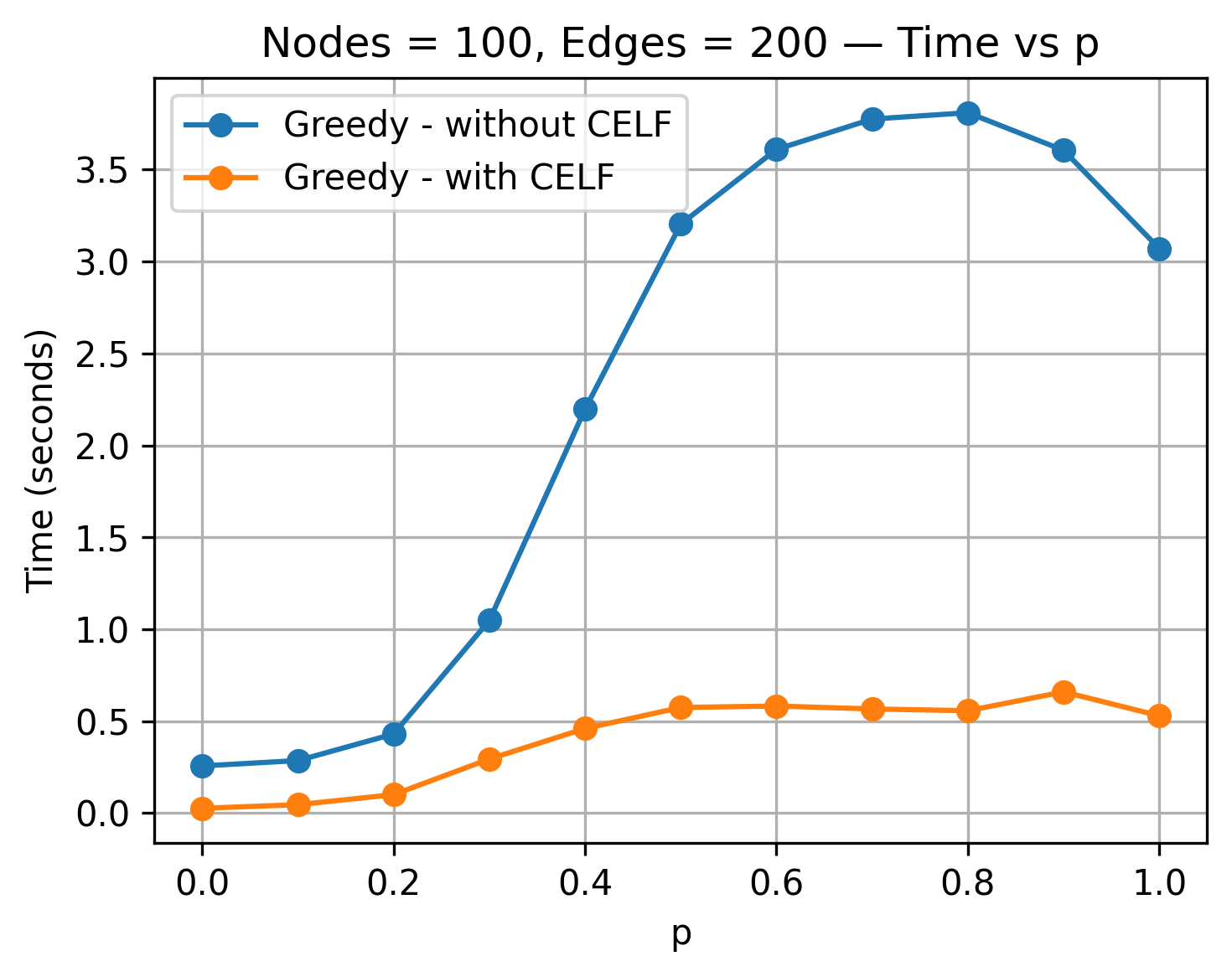}
    \caption{Runtime of Greedy heuristic - with and without CELF optimization}
    \label{fig:csr-comparision}
    \end{figure}

 We integrated the Cost-Effective Lazy Forward (CELF) algorithm introduced by \citet{Leskovec2007CELF} for the influence maximization problem. CELF avoids recalculating the EPC for all nodes in each iteration. In the first iteration, it calculates the EPC for all nodes and sorts them. Then, it only recalculates the spread of the top-ranked node in each subsequent iteration. This significantly speeds up the process without compromising the quality of the solutions found by the algorithm.
 The comparative performance of the greedy algorithm from the empty set variants, with and without CELF, is summarized in Figure~\ref{fig:csr-comparision}.

\subsubsection{Greedy heuristic with MIS}

\begin{figure}[h]
    \centering
    \includegraphics[width=0.6\textwidth]{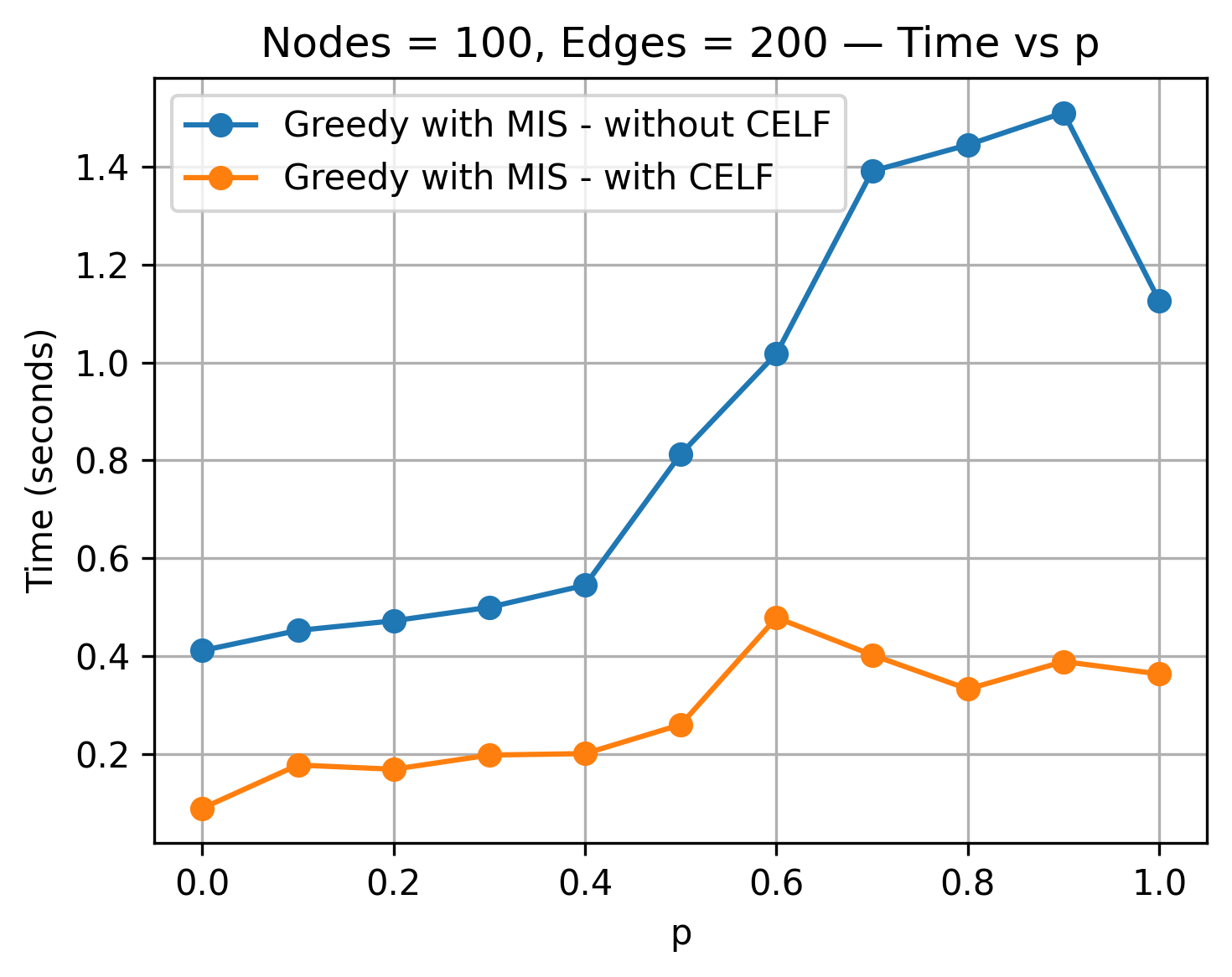}
    \caption{Runtime of Greedy with MIS - with and without CELF optimization}
    \label{fig:csr-comparision-mis}
    \end{figure}

The greedy algorithm with a maximal independent set (MIS) is adapted from the heuristic algorithm of \cite{Arulselvan2009SparseCNDP}, which is originally a deterministic algorithm. We extend it by incorporating the CSR and an EPC estimation algorithm, to make it suitable for stochastic networks. Since MIS generation is inherently random, multiple MIS samples are generated, and the set yielding the lowest EPC is chosen. To improve robustness, the algorithm is executed a certain number of times, and CELF optimization is integrated.
The comparative performance is summarized in Figure~\ref{fig:csr-comparision-mis}. Finally, the best initial solution set is refined using the same local search procedure in REGA.

\begin{algorithm}
\caption{Greedy heuristic with MIS}
    \textbf{Input: } Stochastic graph $G(V,E, \pi)$, and budget $k$\\
    \textbf{Output: }$S$ 
    \begin{algorithmic}[1]
        \State $MIS \leftarrow MaximalIndepSet(G)$
        \For {$i=1,\ldots,|V|-k$}
            \State $v\leftarrow \arg\min\limits_{j\in V\setminus MIS} \sigma(V\setminus \{MIS\cup j\}) - \sigma(V\setminus MIS)$
            \State $MIS\leftarrow MIS\cup v$ 
        \EndFor
        \State $S \leftarrow V\setminus MIS$
    \end{algorithmic}
\end{algorithm}

\subsubsection{Adhoc heuristics.}

\begin{enumerate}
    \item \emph{Betweenness}  adopted from \cite{Dinh2015Vulnerability}, which removes nodes based on their influence scores computed from a random-walk process with a damping factor of 0.85.
    \item \emph{PageRank} adopted from \cite{Dinh2015Vulnerability}, which removes nodes based on their influence scores computed from a random-walk process with a damping factor of 0.85.
    \item \emph{Degree-based Centrality} ranks nodes by their number of edges, assuming high‐degree nodes most affect connectivity, and we use it as a baseline.

    The degree centrality of a vertex $i \in V$ in $G'$ is
    \begin{equation*}    
    C_{D}(i)=\sum\limits_{j\in \delta_i} \pi(i,j),
    \end{equation*}
    where $\delta_i$ is the set of neighbor nodes of $i\in V$, i.e. $\delta_i=\{j\in V|(i,j)\in E\}$.

\end{enumerate}

\subsubsection{Graph Representation Learning.}
\label{subsec:grl}

    % To our knowledge, we are the first to handle SCNDP with a learning‑based model.
    
    In recent years, graph neural networks (GNNs) \cite{peng2021graph} have emerged as a powerful paradigm for graph representation learning. Motivated by the demonstrated effectiveness of GNNs \cite{HamiltonGraphsage,veličković2018graph}, we used the graph sample and aggregated embeddings (GraphSAGE) from \cite{HamiltonGraphsage} and integrated with the custom-designed edge-probability-aware graph attention network (GAT) layer \cite{veličković2018graph}.
    
    Originally, there were no specific datasets for the SCNDP. Therefore, we generate various types of synthetic graphs for training purposes. The labels for these graphs are derived through supervision by applying the REGA algorithm to each training instance. The nodes identified for deletion by REGA are labeled critical, while all other nodes are labeled non-critical under various budgets for critical nodes. Hence, the model is designed to imitate the behavior of REGA while ensuring fast inference.
    
    Our architecture consists of five components: a manual node features, a node encoder, an edge-aware attention layer, a node score decoder, and a training objective. For training, we used curriculum learning and proposed two approaches for inference procedures. Each of them is explained as follows:

    % \paragraph{(i) Manual node features.} 
    % \vspace{1.5em}
    % \noindent\textbf{Manual node features} \\
    \begin{enumerate}

        %%%
        \item \emph{Manual node features.} Each node is represented by an 11-dimensional feature vector. 
        % (degree statistics, ego-network connectivity, multi-hop neighborhood volume, and centrality measures including VoteRank, eigenvector centrality, and k-core number). 
        All features are normalized by dividing each feature by its maximum value within the graph.

        %%%
        \item \emph{Node endoder.} GraphSAGE is used as the base encoder to generate node embeddings by aggregating information from fixed-size neighborhoods. Moreover, the inductive property in the model allows for generalization to unseen graph instances at test time, which is crucial for SCNDP because the test instances may differ from the training instances. Due to the simplicity and computational efficiency of GraphSAGE, the overall model is particularly effective for handling graphs that have different topologies and sizes. 

        We stack \emph{three GraphSAGE} layers with \emph{mean} aggregation and hidden dimensions of 256 with 8 heads. Each layer is followed by \emph{batch normalization} for stabilization and acceleration of the training process and  \emph{ReLU} activation function to introduce nonlinearity, enhancing the model's representational capabilities.

        %%%
        \item \emph{Edge‑aware attention.} To model edge uncertainty, GAT is incorporated, which adaptively weights the contribution of each neighbor. In contrast to fixed aggregation methods, the attention mechanism computes attention coefficients that learn which neighbors are most informative. Thus, multi-head attention improves robustness and representation quality.

        We explicitly inject edge survival probabilities into the attention computation. After the GraphSAGE stack, each node $i$ has an embedding $\vec{h}_i \in \mathbb{R}^{d_{hidden}}$ ($d_{hidden} = 256$). Therefore, in the GAT input layer, $\vec{h}_i$ and $\vec{h}_j$ are input embeddings of nodes $i$ and $j$, respectively,  and $p_{ij}$ is the probability of edge $(i, j)$. The output layer produces updated node features $\vec{h}^{'}_i \in \mathbb{R}^{d'}$ ($d' = 256$) by applying a shared linear transformation for each node, parametrized by a \textit{weight matrix} $\mathbf{W} \in \mathbb{R}^{d' \times d_{hidden}}$. The attention coefficient from $i$ to $j$ is computed as follows: 
    
        \begin{equation*}
            e_{ij} = a(
            \mathbf{W}\, \vec{h}_i,
            \mathbf{W}\, \vec{h}_j,
            \,p_{ij})
        \end{equation*}
    
        Here, $a$ is a shared attentional mechanism. Afterward, normalization is applied using the softmax function:
    
        \begin{equation*}
            \alpha_{ij} = softmax_{j}(e_{ij}) 
        \end{equation*}
    
        After normalization, new representations for each node are obtained using these coefficients $\alpha_{ij}$. In this process, node i is reassigned an updated embedding by aggregating information from its neighboring nodes $\mathcal{N}(i)$ using \textit{sigmoid function} $\sigma$.
        
        \begin{equation*}
            \vec{h}_{i}^{'(k)} = \sigma \left( \sum_{j \in \mathcal{N}(i)} \alpha^{(k)}_{ij} \mathbf{W}^{(k)} \vec{h}_{j} \right) 
        \end{equation*}
    
        Similar to \cite{DBLP:journals/corr/VaswaniSPUJGKP17}, GAT uses multi-head attention to aggregate features from multiple heads to improve performance and stabilize the learning process. Furthermore, the final feature representation for node $i$ is obtained by concatenating the outputs from each attention head, where those $K$ heads are executing transformations independently:
        
        % \begin{equation*}
        %     \vec{h}_i' = \mathop{\big\|}_{k=1}^{K} 
        %     \sigma \left( 
        %     \sum_{j \in \mathcal{N}_i} 
        %     \alpha_{ij}^k \mathbf{W}^k \vec{h}_j 
        %     \right)
        % \end{equation*}
    
        \begin{equation*}
            \vec{h}_i' = \mathop{\big\|}_{k=1}^{K} \left( \vec{h}^{'(k)}_{i} \right), 
        \end{equation*}
    
        where $\mathop{\big\|}$ denotes concatenation operation, and $K$ is the number of attention heads. $\vec{h}^{'k}_{i}$ is the output of the $k$-th attention head for node $i$.
    
        In this architecture, injecting edge survival probabilities into the GAT allows the model to explicitly account for edge reliability during the message passing process and emphasize the connection of the neighbor nodes. In addition, the model learns more robust and uncertainty-aware node representations, which leads to better performance and aligns with the stochastic nature of SCNDP.

        %%%
        \item \emph{Node score decoder.} Given the final node embedding $h'_{i}$, we used a single-layer multilayer perceptron, which is a simple feedforward neural network that assign each node $i$ a scalar \emph{criticality score}:
    
        \begin{equation*}
            s_i=\sigma( \mathbf{W}, \vec{h'}_{i})
        \end{equation*}
        
        We interpret $s_i$ as the probability that node $i$ should be removed under the given node deletion budget.

        %%%
        \item \emph{Training objective.} Since our labels are derived from REGA, the critical node detection procedure is framed as a supervised learning setting. Each node has a binary target $y_i \in \{0, 1\}$, the label for node $i$, and $s_i \in (0, 1)$ is the model's predicted critical score. We train the model as a multi-label node classifier by minimizing the average binary cross-entropy loss:
    
        \begin{equation*}
            \mathcal{L} = \frac{1}{|V|} \sum_{i \in V} \mathrm{BCE}(s_i, y_i).   
        \end{equation*}
        
        In implementation, we use PyTorch's \texttt{BCEWithLogitsLoss}, which combines the sigmoid and binary cross-entropy terms in a numerically stable form. We also evaluated a regression task in which the model directly regressed continuous criticality scores using mean squared error, but the binary classification objective consistently produced lower EPC at inference time. Therefore, results are reported for the classification task.
        
    \end{enumerate}

    % \paragraph{(ii) Node encoder.} 
    % \vspace{1.5em}
    % \noindent\textbf{Node encoder} \\

    % \paragraph{(iii) Edge‑aware attention.}
    % \vspace{1.5em}
    % \noindent\textbf{Edge-aware attention} \\

    % \paragraph{(iv) Node score decoder.}
    % \vspace{1.5em}
    % \noindent\textbf{Node score decoder} \\

    % \paragraph{(v)}
    % \vspace{1.5em}
    % \noindent\textbf{Training objective} \\

    % \paragraph{Curriculum learning.}
    % \vspace{1.5em}
    % \noindent\textbf{Curriculum learning} \\
    In the training, we applied the \emph{curriculum learning approach} to improve convergence and handle the scalability issue. The training begins on smaller and simpler graph instances, and gradually includes larger and more complex instances with various probability types to introduce the complexity of datasets to the model progressively. This helps the model adapt to increasing graph size and probability type without destabilizing early training.

    % \paragraph{Inference Procedures.}
    % \vspace{1.5em}
    % \noindent\textbf{Inference Procedures} \\
    Finally, we evaluate two approaches at test time. First approach is the \emph{GNN (1-shot)}. We apply the trained model in a single pass to a test network and then select the top $k$ nodes, where $k$ is the node deletion budget, based on their criticality scores, such as the node ranking procedure. Second approach is the \emph{Greedy GNN.} We employed a greedy procedure that iteratively applies the model to the remaining network. At each step, the node with the highest criticality score is selected and removed, and this process continues until $k$ nodes have been deleted.
    
    \begin{figure}[htbp]
      \centering
      \begin{subfigure}[t]{0.32\textwidth}
        \centering
        \includegraphics[width=\textwidth]{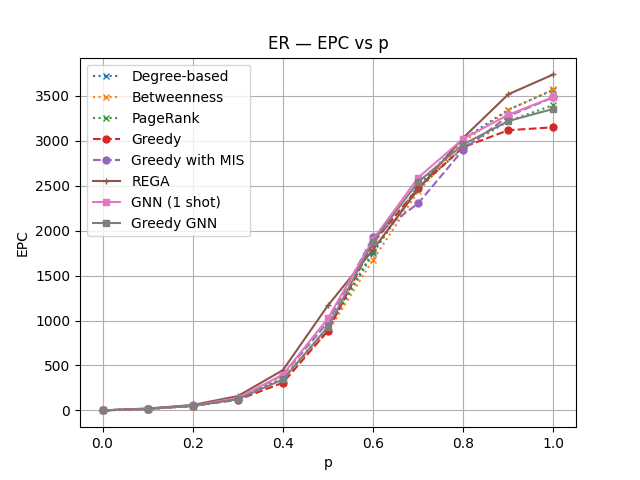}
        \caption{Erdos–Renyi Network}
        \label{fig:er-sparse}
      \end{subfigure}
      \hfill
      \begin{subfigure}[t]{0.32\textwidth}
        \centering
        \includegraphics[width=\textwidth]{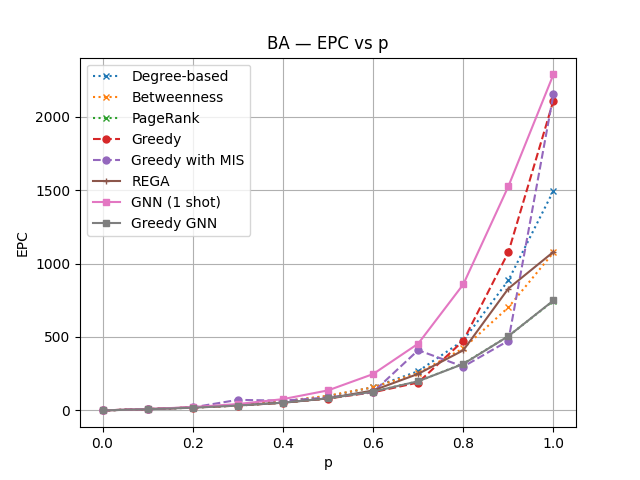}
        \caption{Barabási–Albert Network}
        \label{fig:ba-sparse}
      \end{subfigure}
      \hfill
      \begin{subfigure}[t]{0.32\textwidth}
        \centering
        \includegraphics[width=\textwidth]{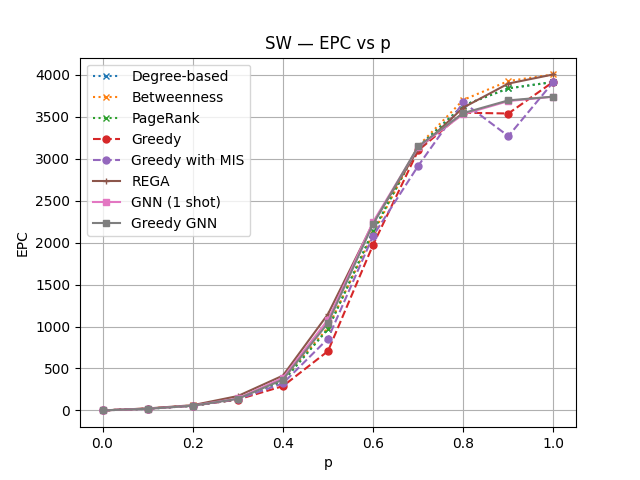}
        \caption{Small-world Network}
        \label{fig:sw-sparse}
      \end{subfigure}

      \caption{Comparing performance of the algorithms: EPC vs p (without Local Search)}
      \label{fig:standard-benchmark-uniform-no-ls-epc}

     \medskip
        
      \begin{subfigure}[t]{0.32\textwidth}
        \centering
        \includegraphics[width=\textwidth]{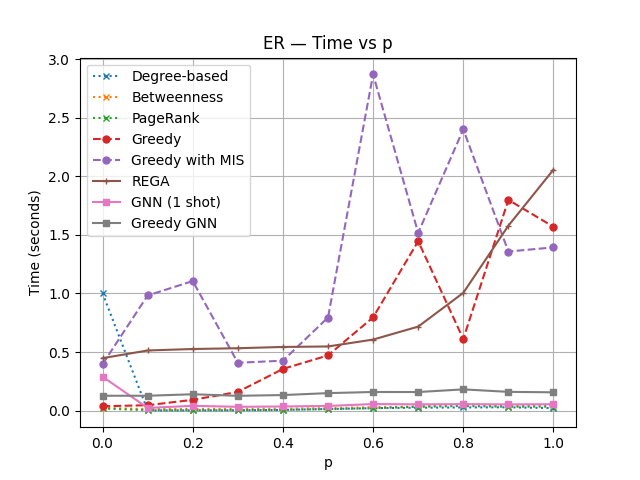}
        \caption{Erdos–Renyi Network}
        \label{fig:er-sparse}
      \end{subfigure}
      \hfill
      \begin{subfigure}[t]{0.32\textwidth}
        \centering
        \includegraphics[width=\textwidth]{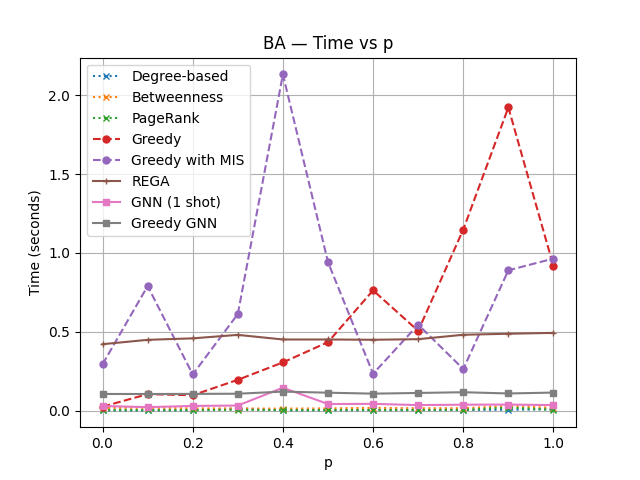}
        \caption{Barabási–Albert Network}
        \label{fig:ba-sparse}
      \end{subfigure}
      \hfill
      \begin{subfigure}[t]{0.32\textwidth}
        \centering
        \includegraphics[width=\textwidth]{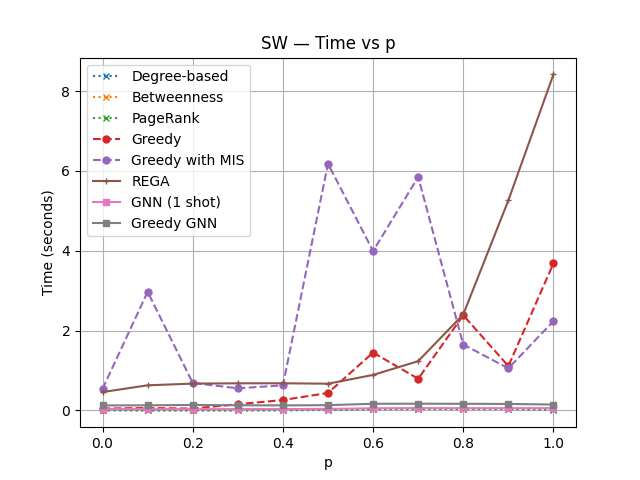}
        \caption{Small-world Network}
        \label{fig:sw-sparse}
      \end{subfigure}

      \caption{Comparing performance of the algorithms: Time vs p (without Local Search)}
      \label{fig:standard-benchmark-uniform-no-ls-runtime}
      \medskip
        
      \begin{subfigure}[t]{0.32\textwidth}
        \centering
        \includegraphics[width=\textwidth]{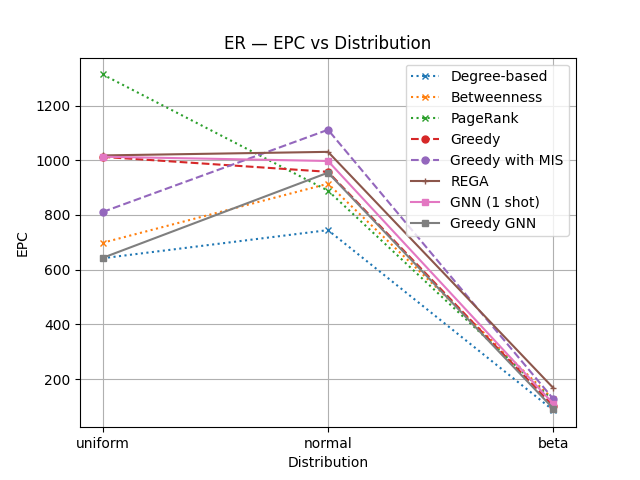}
        \caption{Erdos–Renyi Network}
        \label{fig:er-sparse}
      \end{subfigure}
      \hfill
      \begin{subfigure}[t]{0.32\textwidth}
        \centering
        \includegraphics[width=\textwidth]{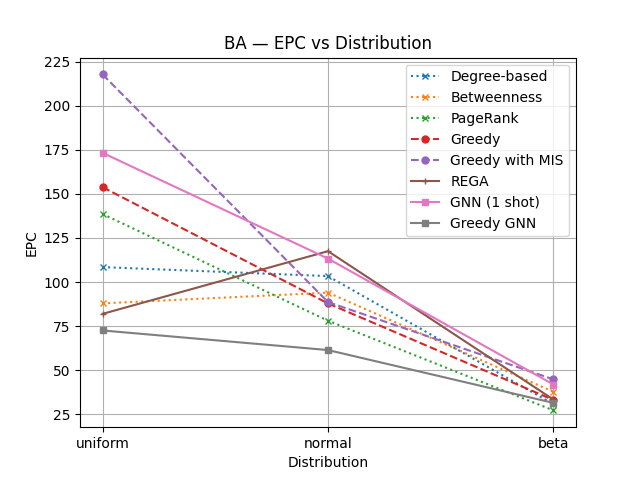}
        \caption{Barabási–Albert Network}
        \label{fig:ba-sparse}
      \end{subfigure}
      \hfill
      \begin{subfigure}[t]{0.32\textwidth}
        \centering
        \includegraphics[width=\textwidth]{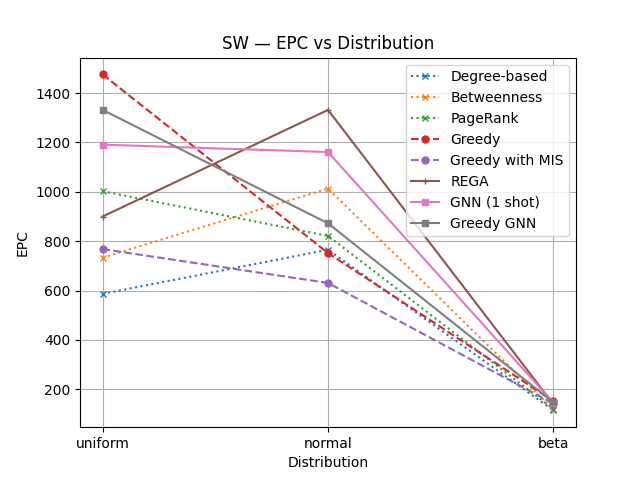}
        \caption{Small-world Network}
        \label{fig:sw-sparse}
      \end{subfigure}
      \caption{Comparing performance of the algorithms (heterogeneous distributions): EPC vs distribution (without Local Search)}
      \label{fig:standard-benchmark-heterogeneous-no-ls-epc}
      
      \medskip
        
      \begin{subfigure}[t]{0.32\textwidth}
        \centering
        \includegraphics[width=\textwidth]{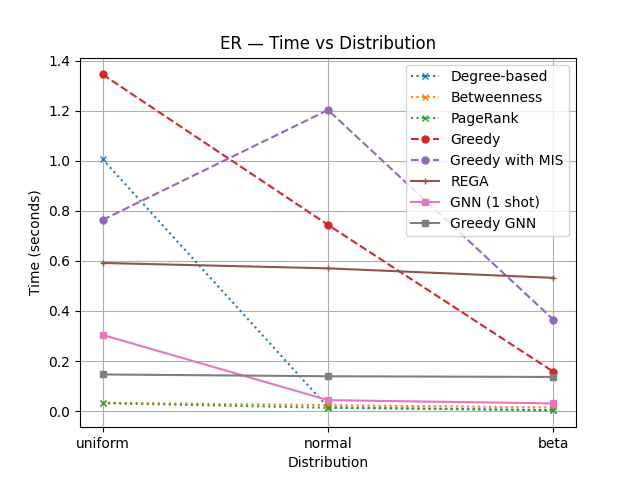}
        \caption{Erdos–Renyi Network}
        \label{fig:er-sparse}
      \end{subfigure}
      \hfill
      \begin{subfigure}[t]{0.32\textwidth}
        \centering
        \includegraphics[width=\textwidth]{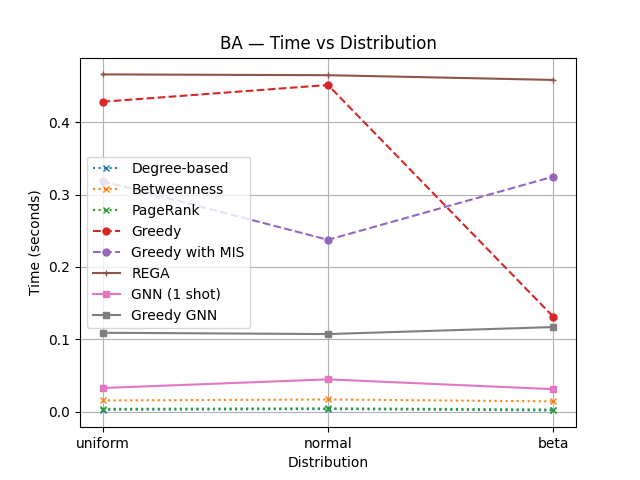}
        \caption{Barabási–Albert Network}
        \label{fig:ba-sparse}
      \end{subfigure}
      \hfill
      \begin{subfigure}[t]{0.32\textwidth}
        \centering
        \includegraphics[width=\textwidth]{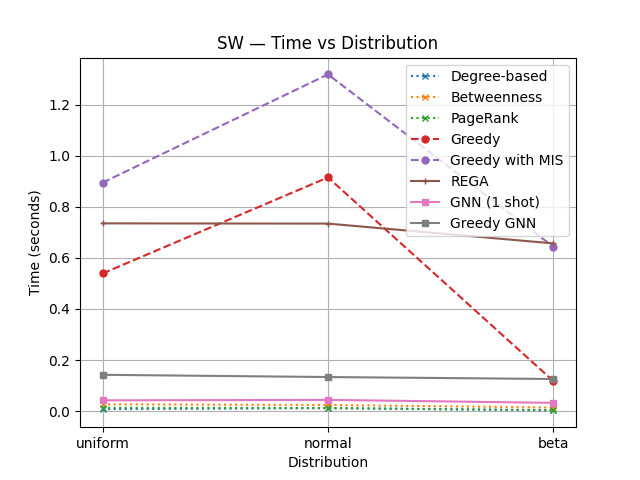}
        \caption{Small-world Network}
        \label{fig:sw-sparse}
      \end{subfigure}
      \caption{Comparing running time of the algorithms (heterogeneous distributions): Time vs distribution (without Local Search)}
      \label{fig:standard-benchmark-heterogeneous-no-ls-runtime}
    \end{figure}
    
\section{Experiments}
\label{sec:experiments}

We demonstrate through experiments the effectiveness of our proposed algorithms in comparison to the REGA heuristic. We evaluated our approaches using randomly generated synthetic graphs that vary in node counts, edge densities, and customized edge probability values or distributions for our experimental settings.
    
\subsection{Edge-probability settings}

We evaluate each proposed method under the following probability settings for edge uncertainty:
\begin{enumerate}
    \item \emph{Uniform probability for all edges}: $\pi(e) = p$ for all $e\in E$, following Dinh and Thai~\cite{Dinh2015Vulnerability}. 
    \item \emph{Heterogeneous probability distributions for all edges:}
    
    % \iffalse
        \[
        \pi(e) \sim 
        \begin{cases}
        \text{Uniform}(0, 1) \\
        \text{Beta}(2, 5) \\
        \text{Normal}(0.5, 0.2)\ \text{truncated to } (0, 1]
        \end{cases}
        \quad \text{for all } e \in E.
        \]
    % \fi
            
\end{enumerate}

\subsection{Dataset}
 To evaluate our methods more comprehensively, we consider a variety of graph instances. We adopt most of the parameters from \cite{Dinh2015Vulnerability} as described below:

 \subsubsection{Standard benchmark datasets}
 \begin{enumerate}
    \item \textit{Erdős–Rényi (ER)}: A random graph of 100 nodes and 200 edges, where each edge is uniformly distributed at random \cite{erdos1960evolution}.
    
    \item \textit{Barabási–Albert (BA)}: A scale-free network of 100 nodes and 200 edges, generated by the preferential attachment mechanism, yields a distribution of degrees of power law \cite{barabasi2000scale}.
    
    \item \textit{Watts–Strogatz (WS)}: A small world network with 100 nodes, derived from a two-dimensional lattice with a rewiring probability of 0.3 \cite{watts1998collective}.
\end{enumerate}

\subsubsection{Larger graph datasets}
    In addition, we generated random graph instances with node counts of 200, 300, and 500 using the same parameters as in the earlier settings. Due to varying node counts, the resulting edge densities differ across random graph types. We evaluated both edge-probability configurations in all test cases.

\subsubsection{Data generation for learning-based approach}
    For the \textit{training and validation dataset}, we generated 720 training and 300 validation graph instances with varying random graph types under the \textit{uniform edge probability} setting. Furthermore, we generated 432 training and 180 validation instances for the \textit{heterogeneous edge probability} setting. The graph instances were generated using different parameters: probabilities of 0.2 and 0.3 for Erdős–Rényi graphs, attachment parameters of 2 and 3 for Barabási–Albert graphs, and rewiring probabilities of 0.3 and 0.4 for Watts–Strogatz graphs.
        % \item  For the \textit{Test dataset}, we generated 120 with node count of 100 and 200 instances with node counts of 200, 300 and 500 with varying random graph types under the \textit{uniform edge probability} setting. Additionally, we created 180 graphs with node count of 100 and 90 instances with node counts of 200, 300 and 500 for the \textit{heterogeneous edge probability} setting.
    
\begin{figure}[htbp]
    \centering
      \begin{subfigure}[t]{0.32\textwidth}
        \centering
        \includegraphics[width=\textwidth]{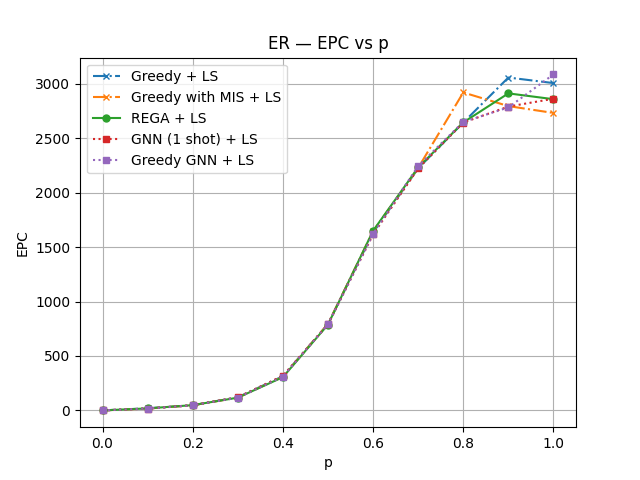}
        \caption{Erdos–Renyi Network}
        \label{fig:er-sparse}
      \end{subfigure}
      \hfill
      \begin{subfigure}[t]{0.32\textwidth}
        \centering
        \includegraphics[width=\textwidth]{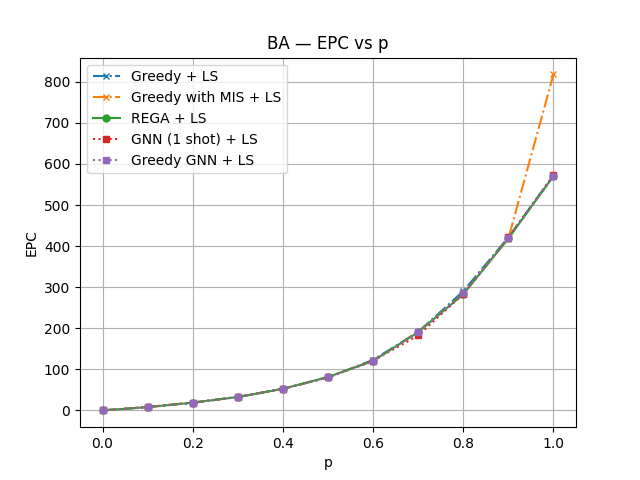}
        \caption{Barabási–Albert Network}
        \label{fig:ba-sparse}
      \end{subfigure}
      \hfill
      \begin{subfigure}[t]{0.32\textwidth}
        \centering
        \includegraphics[width=\textwidth]{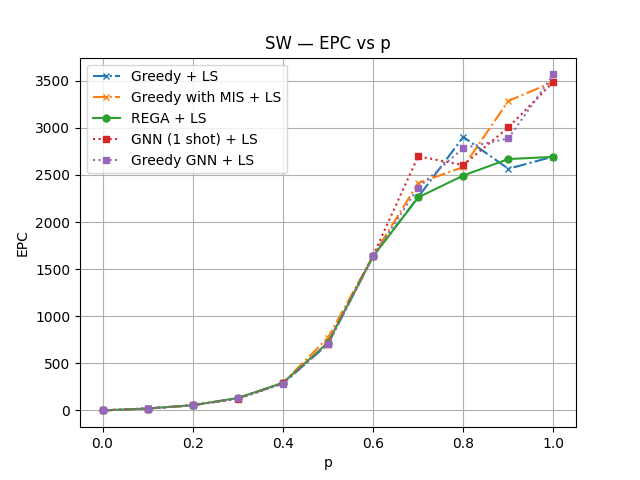}
        \caption{Small-world Network}
        \label{fig:sw-sparse}
      \end{subfigure}

      \caption{Comparing performance of the algorithms: EPC vs p (with local search)}
      \label{fig:standard-benchmark-uniform-with-ls-epc}

     \medskip
        
      \begin{subfigure}[t]{0.32\textwidth}
        \centering
        \includegraphics[width=\textwidth]{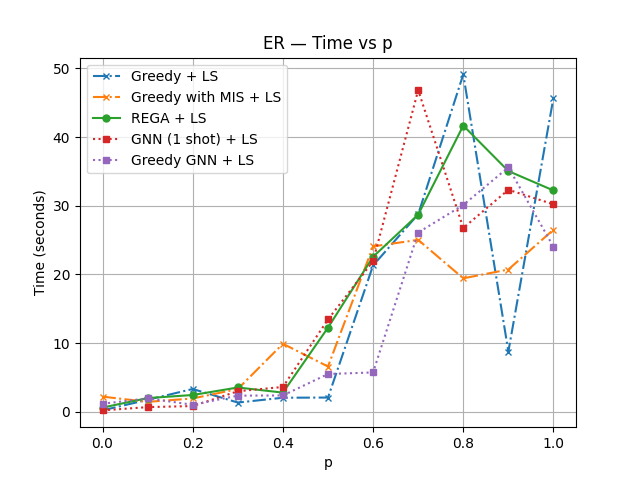}
        \caption{Erdos–Renyi Network}
        \label{fig:er-sparse}
      \end{subfigure}
      \hfill
      \begin{subfigure}[t]{0.32\textwidth}
        \centering
        \includegraphics[width=\textwidth]{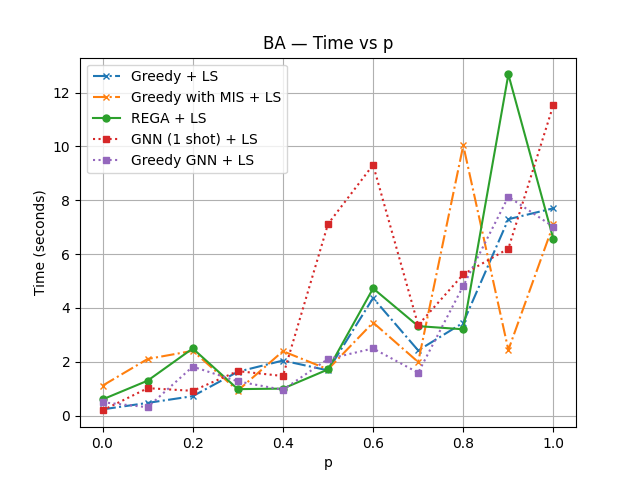}
        \caption{Barabási–Albert Network}
        \label{fig:ba-sparse}
      \end{subfigure}
      \hfill
      \begin{subfigure}[t]{0.32\textwidth}
        \centering
        \includegraphics[width=\textwidth]{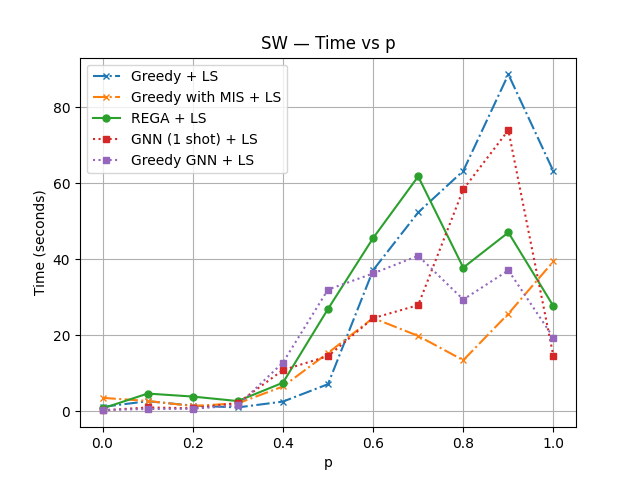}
        \caption{Small-world Network}
        \label{fig:sw-sparse}
      \end{subfigure}
      \caption{Comparing running time of the algorithms: Time vs p (with local search)}
      \label{fig:standard-benchmark-uniform-with-ls-runtime}
      
      \centering
      \begin{subfigure}[t]{0.32\textwidth}
        \centering
        \includegraphics[width=\textwidth]{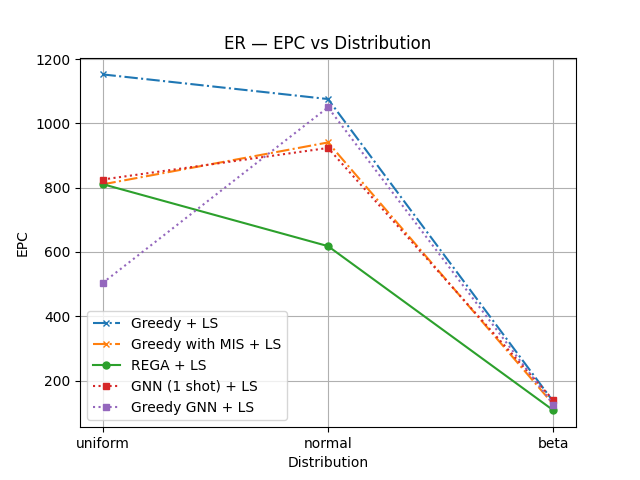}
        \caption{Erdos–Renyi Network}
        \label{fig:er-sparse}
      \end{subfigure}
      \hfill
      \begin{subfigure}[t]{0.32\textwidth}
        \centering
        \includegraphics[width=\textwidth]{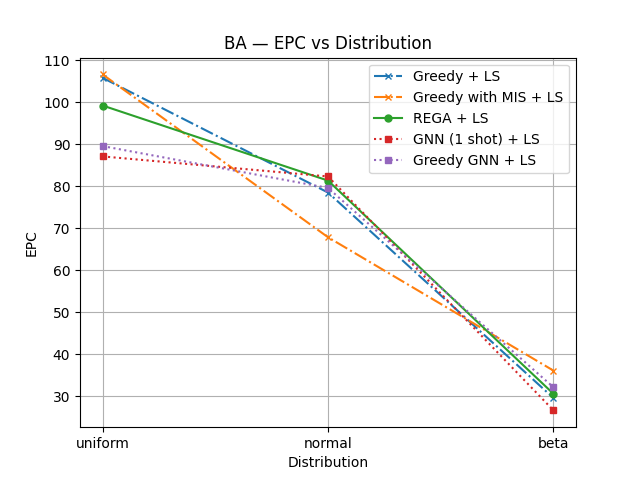}
        \caption{Barabási–Albert Network}
        \label{fig:ba-sparse}
      \end{subfigure}
      \hfill
      \begin{subfigure}[t]{0.32\textwidth}
        \centering
        \includegraphics[width=\textwidth]{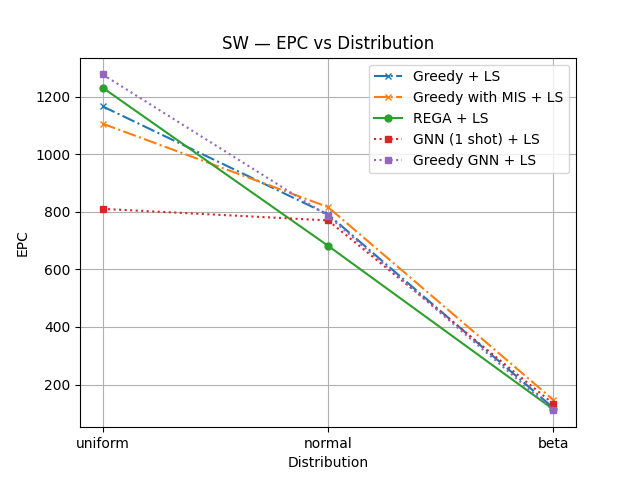}
        \caption{Small-world Network}
        \label{fig:sw-sparse}
      \end{subfigure}

      \caption{Comparing performance of the algorithms (heterogeneous distributions): EPC vs distribution (with local search)}
      \label{fig:standard-benchmark-heterogeneous-with-ls-epc}

     \medskip
        
      \begin{subfigure}[t]{0.32\textwidth}
        \centering
        \includegraphics[width=\textwidth]{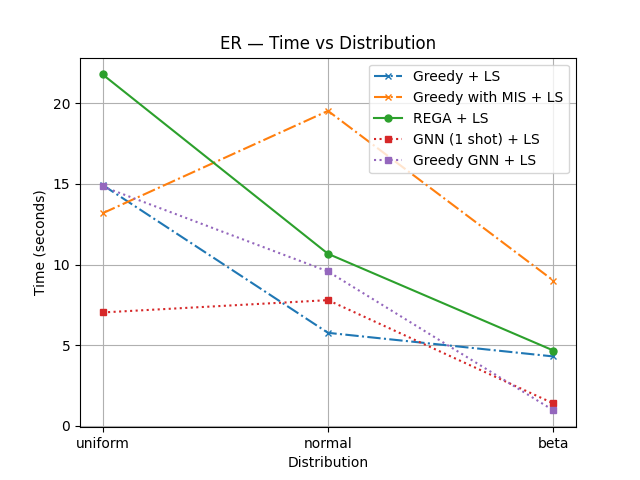}
        \caption{Erdos–Renyi Network}
        \label{fig:er-sparse}
      \end{subfigure}
      \hfill
      \begin{subfigure}[t]{0.32\textwidth}
        \centering
        \includegraphics[width=\textwidth]{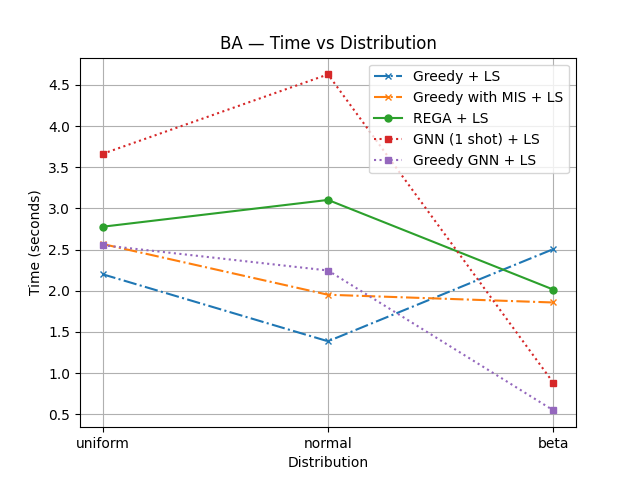}
        \caption{Barabási–Albert Network}
        \label{fig:ba-sparse}
      \end{subfigure}
      \hfill
      \begin{subfigure}[t]{0.32\textwidth}
        \centering
        \includegraphics[width=\textwidth]{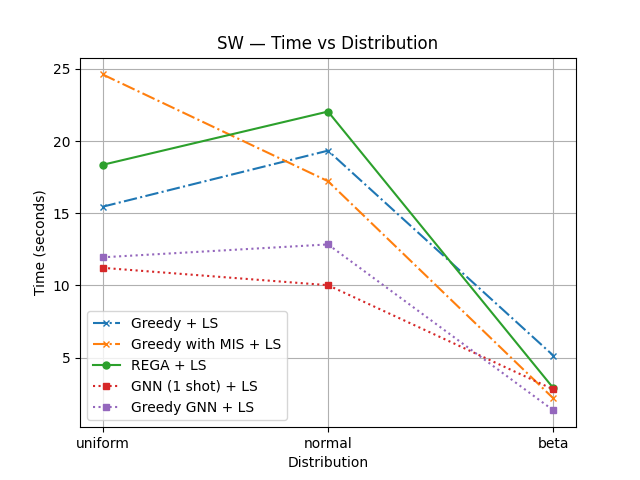}
        \caption{Small-world Network}
        \label{fig:sw-sparse}
      \end{subfigure}
      \caption{Comparing performance of the algorithms (heterogeneous distributions): time vs distribution (with local search)}
      \label{fig:standard-benchmark-heterogeneous-with-ls-runtime}
    \end{figure}
    
\subsection{Settings of Compared Methods}
    In here, we describe the parameter and hyperparameter settings used in the heuristic and learning-based approaches, respectively.
\subsubsection{Heuristic settings.}

    \emph{Greedy with MIS.} Due to the stochastic nature of finding the maximal independent set (MIS), we perform 40 independent trials for graphs with 100 nodes and 20 trials for larger graphs, selecting the MIS yielding the minimum EPC.
\subsubsection{Learning settings.}

    \begin{enumerate}
        \item \textit{Curriculum learning.} We adopt an increasing complexity curriculum, where training begins on smaller graphs with 20, 50, and 80 nodes, and evaluation is carried out on progressively larger instances with 100, 200, 300, and 500 nodes.
        
        \item \textit{Hyper-parameter search.} To find the best hyperparameter configuration, the \textit{Optuna} \cite{10.1145/3292500.3330701} is performed over 50 trials. The best configuration found includes a learning rate of $0.001518$, a weight decay of $10^{-4}$, and the GraphSAGE layer with hidden dimensions of 256, 8 attention heads, and a dropout rate of 0.4. All trials were carried out on an NVIDIA RTX 3080 GPU, and the entire study took approximately 4 hours.
        
        \item \textit{Training and Evaluation.} The model is trained for 30 epochs using the best hyperparameters found by \textit{Optuna}, with a batch size of 128 and the \textit{AdamW optimizer}. Training with \textit{uniform edge probabilities} yields a model that is effectively generalized to both uniform and heterogeneous edge-probability settings. The \textit{ReduceLROnPlateau} learning rate scheduler is used to improve convergence. The loss function used is the \textit{Binary Cross-Entropy loss}. The training time was approximately 20 minutes.
    \end{enumerate}
    
     We use the same 2-exchange local search procedure as employed in \cite{Dinh2015Vulnerability} for all heuristic algorithms and REGA. The local search looks for better solutions by changing one node at a time in the current set and checking to see if the goal is met. We used 10,000 samples for each local search iteration for computational efficiency and low variance in the results. However, 100,000 samples are used to estimate the final EPC, ensuring a highly accurate estimate. 
     
     All methods are evaluated using standard benchmark datasets. The greedy algorithm, the greedy algorithm with Maximum Independent Set (MIS), and the two learning-based approaches are specifically compared on larger graph instances. The results of these comparisons include scenarios both with and without local search procedures. However, local search procedures are not applied to the 300-node and 500-node graphs, since the required computation time grows exponentially with graph size.

\subsection{Environment}
    The algorithms were implemented in Python on a 64-bit Ubuntu platform featuring an Intel i7 3.8 GHz processor and 64 GB of memory. Linear programming problems were solved using the \textit{SciPy} optimization library.

    \begin{table}[htbp]
  \centering
  \caption{EPC and runtime of four algorithms on three network models (Node count = 200, without local search)}
  \label{tab:node-200-no-ls}
  \setlength{\tabcolsep}{2.2pt}
  \small
  \resizebox{\textwidth}{!}{%
  \begin{tabular}{l *{3}{*{4}{c}}}       
    \toprule
        & \multicolumn{4}{c}{\textbf{Erdos-Renyi}}
        & \multicolumn{4}{c}{\textbf{Barabasi-Albert}}
        & \multicolumn{4}{c}{\textbf{Watts-Strogatz}} \\
    \cmidrule(lr){2-5}\cmidrule(lr){6-9}\cmidrule(lr){10-13}
         & Greedy & Greedy MIS & Greedy GNN & GNN (1-shot)
        & Greedy & Greedy MIS & Greedy GNN & GNN (1-shot)
        & Greedy & Greedy MIS & Greedy GNN & GNN (1-shot) \\
    \midrule
    \multicolumn{13}{l}{\textbf{Edge probability \(p\)}}\\
    $p = 0.1$ & 187.8 & \textbf{151.8} & 182.0 & 174.4 & 25.6 & 17.3 & \textbf{16.4} & 18.6 & 43.3 & 39.6 & \textbf{38.4} & 38.7 \\
    \textit{Runtime}       & \textit{0.4} & \textit{1.9} & \textit{1.1} & \textit{0.4} & \textit{0.1} & \textit{2.3} & \textit{0.4} & \textit{0.0} & \textit{0.1} & \textit{0.3} & \textit{0.6} & \textit{0.1} \\
    $p = 0.2$ & 2938.4 & \textbf{2669.1} & 3762.7 & 3590.2 & 44.7 & \textbf{38.7} & 39.1 & 47.4 & 134.3 & 113.4 & \textbf{111.5} & 112.9 \\
    \textit{Runtime}       & \textit{4.1} & \textit{0.7} & \textit{1.1} & \textit{0.1} & \textit{0.8} & \textit{0.1} & \textit{0.4} & \textit{0.0} & \textit{0.2} & \textit{0.2} & \textit{0.6} & \textit{0.0} \\
    $p = 0.3$ & 10677.2 & \textbf{10272.8} & 10619.6 & 10594.1 & 74.4 & \textbf{73.2} & 74.3 & 95.5 & 310.1 & \textbf{272.6} & 285.6 & 287.1 \\
    \textit{Runtime}       & \textit{5.5} & \textit{3.7} & \textit{1.1} & \textit{0.2} & \textit{0.6} & \textit{0.1} & \textit{0.4} & \textit{0.0} & \textit{0.2} & \textit{0.3} & \textit{0.6} & \textit{0.1} \\
    $p = 0.4$ & 14060.6 & \textbf{13457.3} & 13489.8 & 13618.6 & \textbf{118.6} & 118.9 & 132.7 & 182.4 & \textbf{693.5} & 705.7 & 890.9 & 905.3 \\
    \textit{Runtime}       & \textit{8.0} & \textit{1.6} & \textit{1.2} & \textit{0.2} & \textit{1.2} & \textit{0.1} & \textit{0.4} & \textit{0.0} & \textit{1.4} & \textit{0.3} & \textit{0.6} & \textit{0.1} \\
    $p = 0.5$ & 14973.7 & 14890.1 & \textbf{14681.7} & 14923.0 & 178.0 & \textbf{176.0} & 242.6 & 366.0 & 3142.9 & \textbf{2692.5} & 3476.2 & 3623.5 \\
    \textit{Runtime}       & \textit{6.4} & \textit{11.3} & \textit{1.2} & \textit{0.2} & \textit{2.4} & \textit{0.2} & \textit{0.4} & \textit{0.0} & \textit{2.4} & \textit{2.0} & \textit{0.6} & \textit{0.1} \\
    $p = 0.6$ & 15406.3 & 15281.7 & \textbf{15167.4} & 15535.3 & \textbf{282.3} & 319.3 & 477.6 & 825.2 & 9501.9 & \textbf{8145.7} & 8776.7 & 9105.5 \\
    \textit{Runtime}       & \textit{8.5} & \textit{1.6} & \textit{1.3} & \textit{0.2} & \textit{4.8} & \textit{0.2} & \textit{0.4} & \textit{0.1} & \textit{4.4} & \textit{0.6} & \textit{0.6} & \textit{0.1} \\
    $p = 0.7$ & 15490.4 & \textbf{15049.3} & 15411.8 & 15831.8 & 579.3 & \textbf{464.7} & 1005.3 & 1985.4 & 12597.1 & \textbf{12059.9} & 12480.6 & 12775.1 \\
    \textit{Runtime}       & \textit{8.1} & \textit{2.8} & \textit{1.3} & \textit{0.2} & \textit{3.1} & \textit{0.1} & \textit{0.4} & \textit{0.1} & \textit{5.5} & \textit{0.7} & \textit{0.7} & \textit{0.1} \\
    $p = 0.8$ & 15553.9 & 15975.8 & \textbf{15505.9} & 15981.8 & 3084.5 & \textbf{924.6} & 2163.5 & 4116.8 & \textbf{14129.8} & 14358.3 & 14201.2 & 14414.1 \\
    \textit{Runtime}       & \textit{8.1} & \textit{2.4} & \textit{1.2} & \textit{0.2} & \textit{2.2} & \textit{0.2} & \textit{0.4} & \textit{0.1} & \textit{4.0} & \textit{0.7} & \textit{0.6} & \textit{0.1} \\
    $p = 0.9$ & 15730.4 & 15726.9 & \textbf{15540.5} & 16064.8 & 5588.2 & \textbf{1840.1} & 3836.5 & 6166.2 & \textbf{14237.7} & 15192.5 & 15008.2 & 15184.1 \\
    \textit{Runtime}       & \textit{10.2} & \textit{2.1} & \textit{1.2} & \textit{0.2} & \textit{2.1} & \textit{0.2} & \textit{0.4} & \textit{0.1} & \textit{6.0} & \textit{2.1} & \textit{0.6} & \textit{0.1} \\
    $p = 1.0$ & 16110.0 & \textbf{15233.0} & 15569.9 & 16110.0 & 11338.0 & \textbf{3038.0} & 5336.2 & 7411.5 & 15750.6 & 15934.5 & \textbf{15403.5} & 15577.1 \\
    \textit{Runtime}       & \textit{82.8} & \textit{3.4} & \textit{1.2} & \textit{0.2} & \textit{2.5} & \textit{0.2} & \textit{0.4} & \textit{0.1} & \textit{43.3} & \textit{0.6} & \textit{0.6} & \textit{0.1} \\
    
    \addlinespace
    
    \multicolumn{13}{l}{\textbf{Edge-probability distribution}}\\
    Uniform & 15123.7 & 15030.9 & \textbf{14993.8} & 15013.1 & \textbf{248.8} & 257.8 & 257.6 & 331.4 & 3195.5 & \textbf{2868.2} & 4380.1 & 3912.4 \\
    \textit{Runtime}       & \textit{15.3} & \textit{8.3} & \textit{1.2} & \textit{0.7} & \textit{4.7} & \textit{0.6} & \textit{0.4} & \textit{0.1} & \textit{4.3} & \textit{0.3} & \textit{0.6} & \textit{0.1} \\
    Beta & 9824.1 & 10030.5 & 10123.7 & \textbf{9781.8} & 76.0 & 72.0 & \textbf{64.6} & 71.8 & 286.6 & 286.3 & \textbf{239.7} & 313.7 \\
    \textit{Runtime}       & \textit{5.0} & \textit{2.5} & \textit{1.1} & \textit{0.2} & \textit{0.6} & \textit{0.1} & \textit{0.4} & \textit{0.1} & \textit{1.3} & \textit{0.2} & \textit{0.7} & \textit{0.1} \\
    Normal & 14983.2 & 15156.5 & \textbf{14650.2} & 14820.2 & \textbf{200.4} & 250.5 & 241.3 & 343.5 & \textbf{2881.0} & 2986.2 & 3961.4 & 4726.6 \\
    \textit{Runtime}       & \textit{9.9} & \textit{3.6} & \textit{1.3} & \textit{0.2} & \textit{3.0} & \textit{0.2} & \textit{0.4} & \textit{0.1} & \textit{4.2} & \textit{8.3} & \textit{0.7} & \textit{0.1} \\
    \bottomrule
  \end{tabular}}
\end{table}

\subsection{Experimental Results}
    To evaluate our proposed methods, we conducted two groups of experiments. In the first group, we report all of the results of the proposed method on 100-node standard benchmark datasets. In the second group, we employed only selected methods for larger 200, 300, and 500-node graphs because of manageable processing time. Then, their results are improved through local searches. We experiment under both uniform probability and heterogeneous probability distributions for all edges.
    
\subsubsection{Standard benchmark instances.}
    Across algorithms, lower EPC values and shorter runtimes indicate the best performance.

    \begin{enumerate}
        \item \emph{Results without local search procedure.} Figures \ref{fig:standard-benchmark-uniform-no-ls-epc} and \ref{fig:standard-benchmark-heterogeneous-no-ls-epc} indicate the experimental results obtained with the uniform probabilities and with the heterogeneous probabilities for all edges, respectively. The naive REGA algorithm, which does not incorporate a local search procedure, produced lower-quality solutions and had low efficiency. In general, ad hoc heuristics perform relatively better than other algorithms under both probability settings. Furthermore, Greedy GNN and Greedy with MIS achieved lower EPC values in most cases, whereas Greedy GNN shows more stable performance, which emphasizes the benefit of an injection of edge probability. In contrast, the greedy heuristic and the GNN (1-shot) illustrate moderate performance, achieving competitive results in certain network types while falling short in others.
    
        The runtimes of the algorithms are illustrated in \ref{fig:standard-benchmark-uniform-no-ls-runtime} and \ref{fig:standard-benchmark-heterogeneous-no-ls-runtime}. The runtimes for two learning-based approaches and adhoc heuristics are significantly low and consistent across all types of networks. Particularly for learning-based approaches, they benefit from one-shot inference and the elimination of iterative search or recomputation. Other algorithms, including Greedy, Greedy with MIS, and REGA, encounter a slight runtime growth as edge probability increases, but their overall time consumption remains very low.
        
        \item \emph{Results with local search procedure.} Solution quality improved significantly when integrated with the local search procedure but resulted in a considerably increased runtime for all algorithms, illustrated in figures \ref{fig:standard-benchmark-uniform-with-ls-epc} and \ref{fig:standard-benchmark-heterogeneous-with-ls-epc}. Adhoc heuristics are excluded from comparison since their performance was poor. In the edge uniform probability setting, all algorithms yield nearly identical EPC values for the first several edge probability values. Thus, REGA significantly benefited from local search, resulting in lower EPC values. Methods such as Greedy, Greedy-MIS, and Greedy GNN also achieved lower EPC values, often outperforming REGA in terms of overall effectiveness. In contrast, the GNN (1-shot) performs adequately, showing improvements compared to its earlier results without a local search procedure. Nonetheless, Greedy GNN shows reduced effectiveness in a heterogeneous probability setting.

        In terms of runtime, the local search procedure significantly increases the runtime of all algorithms, which are shown in \ref{fig:standard-benchmark-uniform-with-ls-runtime} and \ref{fig:standard-benchmark-heterogeneous-with-ls-runtime}, when compared to the earlier results. In contrast, other algorithms show significantly lower runtime and exhibit considerable fluctuations. REGA maintains a moderate runtime, effectively balancing it with solution quality, whereas the Greedy, the Greedy with MIS, and the two learning-based approaches illustrate efficient computation in both probability settings. 
    \end{enumerate}
    
    Overall, our proposed algorithms offer comparable solution quality to REGA, both with and without the local search procedure. Therefore, it significantly reduces computational cost, making it particularly suitable for larger graph instances.
    
\subsubsection{Larger graph instances.}

    The tables \ref{tab:node-200-no-ls}, \ref{tab:node-200-with-ls}, \ref{tab:node-300-no-ls} and \ref{tab:node-500-no-ls} illustrate the experimental results of the Greedy, Greedy with MIS, Greedy GNN, and GNN (1-shot) on large graph instances. The consumption time of the local search procedure on larger graphs, including those with 200, 300, and 500 nodes, was massive. Therefore, we only applied the procedure to 200-node graphs. 
    
    In the 200-node graph experiments, illustrated in the tables \ref{tab:node-200-no-ls} and \ref{tab:node-200-with-ls}, the runtime of the greedy algorithm increases notably with higher per-edge probabilities, reaching its peak when $p = 1.0$. On the other hand, GNN's (1-shot) runtime consistently achieves the lowest among all methods, demonstrating the efficiency of the learning-based approach. The Greedy has a slightly good performance in the BA and SW random graph, but in general, Greedy and GNN (1-shot) methods achieve weak performance in EPC values, whereas Greedy with MIS and Greedy GNN show the best performance in a trade-off between EPC and runtime, showing competitive EPC values while maintaining manageable computational cost. When integrated with the local search procedure, the runtime of the greedy algorithm wins in a few cases, particularly in ER networks, although its overall performance was worse. Conversely, Greedy with MIS and Greedy GNN continue to leverage both good solution quality and runtime efficiency.

    In 300 and 500-node graph experimental results, illustrated in the tables \ref{tab:node-300-no-ls} and \ref{tab:node-500-no-ls}, the Greedy with MIS and the Greedy GNN maintain the best performance trade-off between EPC values and runtime, but Greedy GNN performs poorly, especially in ER graphs with 300 nodes, while Greedy wins over it in certain times. Furthermore, EPC values in all algorithms tend to converge to a constant value, particularly in ER and SW random graph models, when the number of nodes, edges, and per-edge probability increases, reflecting the saturation of network connectivity. Consequently, Greedy with MIS and the learning-based approaches continue to achieve promising EPC results under manageable runtimes, demonstrating their scalability to larger random graphs.  
    
\subsubsection{Impact of local search procedure.}
    The integration of local search significantly improved EPC outcomes across all methods in both uniform and heterogeneous edge-probability settings. The Greedy and Greedy with MIS methods demonstrate moderate gains, indicating a moderate reliance on the local search procedure. In contrast, REGA depends significantly on the local search procedure to achieve acceptable performance. As a result, the learning-based approaches illustrate varying levels of reliance on local search procedures, ranging from minimal to substantial.

    \begin{figure}[h]
    \centering
    \includegraphics[width=1.0\textwidth]{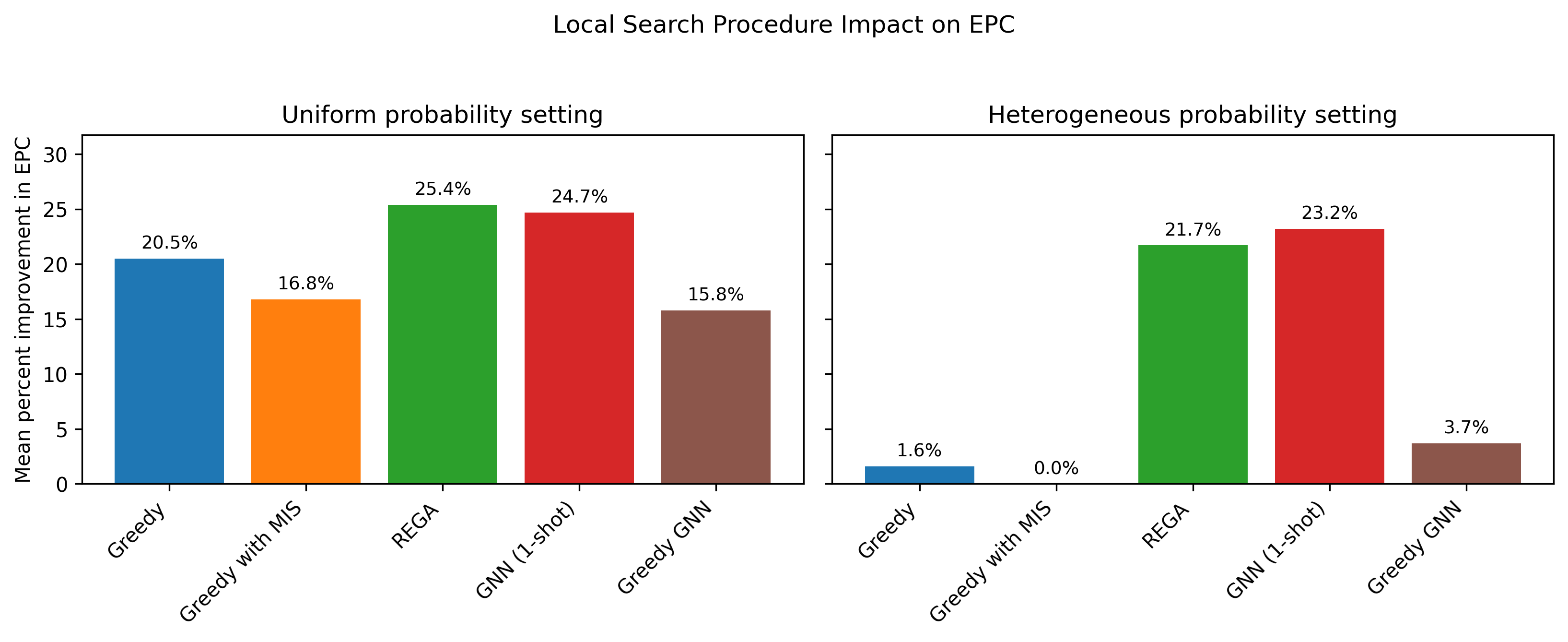}
    \caption{Impact of Local Search Procedure}
    \label{fig:csr-compare}
    \end{figure}

% \newpage
\section{Conclusion}
\label{sec:conclusion}
    In this study, we proposed heuristics and learning-based approaches for SCNDP, namely, Greedy with MIS, and GNN-based architecture methods. In the benchmark evaluation, synthetic random graphs, including ER, SW, and BA, are used with varying node counts, edge densities, and different topologies, where edge survival was drawn from uniform and heterogeneous distributions. We compared the proposed algorithms with REGA. For a consistent experiment, the local search procedure was used for each algorithm to refine the results. The results indicate heuristic approaches achieve equal or lower EPC in certain cases and demonstrate faster runtime, while learning-based approaches were the fastest and showed the potential of imitation learning by better handling the scalability issue due to the near-constant inference time. Future research will involve integrating more complex architectures and deep reinforcement learning into the learning-based approach and assessing the algorithm’s performance on much larger and real-life graphs.

\newpage
    \begin{table}[htbp]
        \centering
        \caption{EPC and runtime of four algorithms on three network models (Node count = 200, with local search)}
        \label{tab:node-200-with-ls}
        \setlength{\tabcolsep}{2.2pt}
        \small
        \resizebox{\textwidth}{!}{%
        \begin{tabular}{l *{3}{*{4}{c}}}       
        \toprule
          & \multicolumn{4}{c}{\textbf{Erdos-Renyi}}
          & \multicolumn{4}{c}{\textbf{Barabasi-Albert}}
          & \multicolumn{4}{c}{\textbf{Watts-Strogatz}} \\
        \cmidrule(lr){2-5}\cmidrule(lr){6-9}\cmidrule(lr){10-13}
           & Greedy & Greedy MIS & Greedy GNN & GNN (1-shot)
          & Greedy & Greedy MIS & Greedy GNN & GNN (1-shot)
          & Greedy & Greedy MIS & Greedy GNN & GNN (1-shot) \\
        \midrule
        \multicolumn{13}{l}{\textbf{Edge probability \(p\)}}\\
        $p = 0.1$ & 171.0 & \textbf{151.5} & 183.8 & 154.9 & 18.5 & 17.3 & \textbf{15.9} & 17.7 & 41.8 & 38.8 & \textbf{38.3} & 39.1 \\
        \textit{Runtime} & \textit{59.6} & \textit{33.3} & \textit{12.4} & \textit{57.3} & \textit{15.3} & \textit{10.4} & \textit{5.8} & \textit{8.4} & \textit{13.8} & \textit{11.8} & \textit{5.8} & \textit{8.1} \\
        $p = 0.2$ & 2607.4 & 2628.3 & \textbf{2575.9} & 2648.0 & 38.5 & 38.8 & \textbf{38.1} & 39.9 & 120.9 & 113.5 & \textbf{110.7} & 113.7 \\
        \textit{Runtime} & \textit{187.3} & \textit{109.8} & \textit{363.8} & \textit{241.4} & \textit{19.7} & \textit{14.7} & \textit{4.7} & \textit{28.2} & \textit{13.5} & \textit{12.1} & \textit{18.0} & \textit{3.7} \\
        $p = 0.3$ & 10210.1 & 10186.8 & 10377.1 & \textbf{10181.9} & 68.2 & 69.3 & \textbf{67.6} & 73.1 & \textbf{257.6} & 271.0 & 258.4 & 258.8 \\
        \textit{Runtime} & \textit{982.3} & \textit{289.9} & \textit{360.2} & \textit{477.3} & \textit{9.8} & \textit{16.7} & \textit{16.2} & \textit{20.5} & \textit{26.1} & \textit{14.9} & \textit{38.2} & \textit{30.5} \\
        $p = 0.4$ & 13584.2 & 13443.0 & \textbf{13442.1} & 13491.0 & 110.7 & \textbf{110.4} & 112.3 & 124.5 & \textbf{700.7} & 732.9 & 709.3 & 720.8 \\
        \textit{Runtime} & \textit{261.3} & \textit{233.7} & \textit{130.2} & \textit{456.1} & \textit{17.0} & \textit{12.8} & \textit{31.3} & \textit{15.5} & \textit{18.7} & \textit{64.1} & \textit{48.8} & \textit{66.9} \\
        $p = 0.5$ & 14868.3 & 14611.1 & \textbf{14505.6} & 14519.6 & 177.7 & 179.3 & 174.8 & \textbf{174.3} & \textbf{2388.8} & 2585.9 & 2622.1 & 2416.9 \\
        \textit{Runtime} & \textit{253.5} & \textit{538.0} & \textit{418.9} & \textit{596.2} & \textit{14.3} & \textit{23.5} & \textit{18.9} & \textit{48.2} & \textit{100.1} & \textit{154.9} & \textit{82.5} & \textit{237.0} \\
        $p = 0.6$ & \textbf{14726.3} & 14955.4 & 14855.2 & 14917.5 & 279.8 & \textbf{279.2} & 282.0 & 297.6 & 7606.5 & \textbf{7440.8} & 7817.9 & 7698.9 \\
        \textit{Runtime} & \textit{733.4} & \textit{484.1} & \textit{357.1} & \textit{538.3} & \textit{9.2} & \textit{38.2} & \textit{27.4} & \textit{40.7} & \textit{443.0} & \textit{421.7} & \textit{295.4} & \textit{412.8} \\
        $p = 0.7$ & \textbf{14942.0} & 15142.9 & 14979.4 & 15140.9 & 452.0 & 452.1 & \textbf{449.0} & 451.2 & 10971.3 & \textbf{10295.3} & 10297.7 & 10297.3 \\
        \textit{Runtime} & \textit{623.9} & \textit{374.5} & \textit{417.4} & \textit{598.2} & \textit{12.4} & \textit{48.9} & \textit{35.5} & \textit{93.1} & \textit{542.9} & \textit{366.8} & \textit{728.5} & \textit{638.7} \\
        $p = 0.8$ & 15198.2 & 15024.5 & 15041.4 & \textbf{15021.1} & 784.9 & 926.7 & \textbf{740.7} & 741.0 & 12394.2 & \textbf{11703.8} & 11727.4 & 12344.2 \\
        \textit{Runtime} & \textit{626.9} & \textit{451.1} & \textit{380.1} & \textit{794.6} & \textit{86.1} & \textit{72.4} & \textit{49.1} & \textit{98.1} & \textit{479.7} & \textit{496.0} & \textit{774.1} & \textit{697.8} \\
        $p = 0.9$ & 15746.8 & 15397.5 & 15391.4 & \textbf{15058.9} & 1265.2 & 1335.9 & 1265.5 & \textbf{1259.4} & 11638.6 & 12091.2 & \textbf{11649.1} & 12473.7 \\
        \textit{Runtime} & \textit{146.3} & \textit{196.6} & \textit{189.6} & \textit{784.5} & \textit{162.0} & \textit{69.2} & \textit{95.1} & \textit{127.8} & \textit{880.3} & \textit{472.5} & \textit{764.9} & \textit{643.3} \\
        $p = 1.0$ & 16110.0 & 15749.8 & \textbf{15398.0} & 15758.3 & 3003.5 & 2203.5 & \textbf{2095.6} & 4281.1 & \textbf{12423.6} & 14879.2 & 12824.8 & 13373.6 \\
        \textit{Runtime} & \textit{163.9} & \textit{190.5} & \textit{317.1} & \textit{253.4} & \textit{279.4} & \textit{26.6} & \textit{157.5} & \textit{153.1} & \textit{824.3} & \textit{420.0} & \textit{536.6} & \textit{409.6} \\
        
        \addlinespace
        \multicolumn{13}{l}{\textbf{Edge-probability distribution}}\\
        Uniform & 14935.3 & 14822.4 & \textbf{14610.6} & 14933.5 & 199.5 & 229.2 & 204.6 & \textbf{159.7} & 4312.7 & \textbf{3040.0} & 3321.2 & 3157.6 \\
        \textit{Runtime} & \textit{785.2} & \textit{1045.8} & \textit{619.6} & \textit{851.6} & \textit{43.6} & \textit{55.0} & \textit{27.7} & \textit{28.9} & \textit{168.3} & \textit{269.0} & \textit{120.4} & \textit{151.7} \\
        Beta & 9638.5 & 9906.6 & 9907.3 & \textbf{9683.2} & 68.3 & 67.0 & 76.1 & \textbf{63.4} & 258.6 & 259.0 & \textbf{233.5} & 240.2 \\
        \textit{Runtime} & \textit{709.9} & \textit{881.8} & \textit{921.0} & \textit{722.7} & \textit{37.8} & \textit{20.8} & \textit{11.8} & \textit{12.9} & \textit{55.3} & \textit{37.2} & \textit{50.1} & \textit{40.9} \\
        Normal & 14726.1 & 15109.3 & \textbf{14608.5} & 14723.7 & \textbf{163.3} & 214.9 & 171.5 & 193.7 & 3953.7 & \textbf{1610.6} & 3133.1 & 2897.9 \\
        \textit{Runtime} & \textit{1265.8} & \textit{306.4} & \textit{498.8} & \textit{754.4} & \textit{8.5} & \textit{37.0} & \textit{15.1} & \textit{36.9} & \textit{169.4} & \textit{231.4} & \textit{134.3} & \textit{132.2} \\
        \bottomrule
        \end{tabular}}
    \end{table}

    \begin{table}[htbp]
        \centering
        \caption{EPC and runtime of four algorithms on three network models (node count = 300, without local search)}
        \label{tab:node-300-no-ls}
        \setlength{\tabcolsep}{2.2pt}
        \small
        \resizebox{\textwidth}{!}{%
        \begin{tabular}{l *{3}{*{4}{c}}}        
        \toprule
        & \multicolumn{4}{c}{\textbf{Erdos-Renyi}}
        & \multicolumn{4}{c}{\textbf{Barabasi-Albert}}
        & \multicolumn{4}{c}{\textbf{Watts-Strogatz}} \\
        \cmidrule(lr){2-5}\cmidrule(lr){6-9}\cmidrule(lr){10-13}
        & Greedy & Greedy MIS & Greedy GNN & GNN (1-shot)
        & Greedy & Greedy MIS & Greedy GNN & GNN (1-shot)
        & Greedy & Greedy MIS & Greedy GNN & GNN (1-shot) \\
        \midrule
        \multicolumn{13}{l}{\textbf{Edge probability \(p\)}}\\
        $p = 0.1$ & 1855.9 & \textbf{1578.0} & 3057.4 & 2424.6 & 41.1 & 29.3 & \textbf{26.6} & 35.8 & 67.2 & 61.7 & \textbf{58.0} & 57.9 \\
        \textit{Runtime} & \textit{5.8} & \textit{3.5} & \textit{4.7} & \textit{0.5} & \textit{0.4} & \textit{0.3} & \textit{1.1} & \textit{0.1} & \textit{0.4} & \textit{1.1} & \textit{1.5} & \textit{0.1} \\
        $p = 0.2$ & 27694.9 & \textbf{25607.2} & 27290.8 & 25797.8 & 106.9 & 68.6 & \textbf{65.3} & 100.8 & 206.6 & 172.5 & \textbf{167.3} & 168.6 \\
        \textit{Runtime} & \textit{21.7} & \textit{10.1} & \textit{5.0} & \textit{0.4} & \textit{0.8} & \textit{1.3} & \textit{1.1} & \textit{0.1} & \textit{2.8} & \textit{1.4} & \textit{1.6} & \textit{0.1} \\
        $p = 0.3$ & 34080.3 & 33178.9 & 33887.2 & \textbf{33016.2} & 144.2 & 136.6 & \textbf{124.5} & 237.5 & 586.6 & 428.7 & \textbf{424.5} & 436.3 \\
        \textit{Runtime} & \textit{29.4} & \textit{7.0} & \textit{5.2} & \textit{0.5} & \textit{2.0} & \textit{0.3} & \textit{1.1} & \textit{0.1} & \textit{0.6} & \textit{0.5} & \textit{1.6} & \textit{0.1} \\
        $p = 0.4$ & 35556.1 & 35325.5 & 35509.8 & \textbf{35046.1} & \textbf{193.4} & 342.7 & 231.0 & 615.7 & \textbf{1163.2} & 1284.1 & 1371.6 & 1434.5 \\
        \textit{Runtime} & \textit{27.9} & \textit{9.4} & \textit{5.2} & \textit{0.5} & \textit{3.8} & \textit{0.5} & \textit{1.1} & \textit{0.1} & \textit{2.0} & \textit{0.8} & \textit{1.6} & \textit{0.2} \\
        $p = 0.5$ & 36095.3 & \textbf{35661.7} & 35972.9 & 35675.9 & \textbf{324.7} & 347.4 & 438.9 & 1874.3 & 7841.9 & \textbf{5513.5} & 6923.3 & 7074.5 \\
        \textit{Runtime} & \textit{26.8} & \textit{13.4} & \textit{5.2} & \textit{0.6} & \textit{4.4} & \textit{0.3} & \textit{1.1} & \textit{0.1} & \textit{8.0} & \textit{1.6} & \textit{1.6} & \textit{0.1} \\
        $p = 0.6$ & \textbf{35800.1} & 35968.4 & 36134.5 & 35933.5 & \textbf{516.8} & 772.3 & 924.1 & 5476.2 & \textbf{23300.2} & 19290.8 & 20307.8 & 19824.7 \\
        \textit{Runtime} & \textit{46.5} & \textit{13.2} & \textit{5.2} & \textit{0.6} & \textit{5.9} & \textit{0.5} & \textit{1.1} & \textit{0.1} & \textit{10.3} & \textit{3.7} & \textit{1.6} & \textit{0.1} \\
        $p = 0.7$ & \textbf{35508.7} & 35942.1 & 36196.9 & 36048.2 & 1507.6 & \textbf{1087.1} & 2204.6 & 10847.5 & 31177.1 & \textbf{28304.6} & 29372.8 & 28322.6 \\
        \textit{Runtime} & \textit{43.5} & \textit{5.8} & \textit{5.1} & \textit{0.5} & \textit{5.1} & \textit{0.3} & \textit{1.1} & \textit{0.1} & \textit{9.5} & \textit{1.5} & \textit{1.6} & \textit{0.2} \\
        $p = 0.8$ & \textbf{36037.4} & 36189.9 & 36248.7 & 36146.2 & 8743.3 & \textbf{2618.2} & 5173.2 & 15444.2 & 33504.0 & 32760.4 & 33673.0 & \textbf{32642.6} \\
        \textit{Runtime} & \textit{48.0} & \textit{6.8} & \textit{5.0} & \textit{0.5} & \textit{5.7} & \textit{1.5} & \textit{1.1} & \textit{0.1} & \textit{9.9} & \textit{2.5} & \textit{1.7} & \textit{0.2} \\
        $p = 0.9$ & \textbf{36047.5} & 36235.5 & 36284.9 & 36235.2 & 15926.8 & \textbf{2586.9} & 9787.2 & 19131.5 & 34775.0 & 34988.7 & 35584.8 & \textbf{34840.5} \\
        \textit{Runtime} & \textit{47.8} & \textit{7.5} & \textit{5.0} & \textit{0.5} & \textit{5.4} & \textit{3.9} & \textit{1.1} & \textit{0.1} & \textit{11.2} & \textit{3.6} & \textit{1.7} & \textit{0.2} \\
        $p = 1.0$ & 36315.0 & \textbf{35784.3} & 36315.0 & 36315.0 & 22583.8 & \textbf{4233.5} & 13949.2 & 21932.7 & 36044.7 & \textbf{35763.8} & 36315.0 & 35778.6 \\
        \textit{Runtime} & \textit{299.6} & \textit{11.1} & \textit{4.8} & \textit{0.5} & \textit{6.1} & \textit{0.6} & \textit{1.1} & \textit{0.1} & \textit{89.6} & \textit{4.1} & \textit{1.6} & \textit{0.1} \\
        
        \addlinespace
        \multicolumn{13}{l}{\textbf{Edge-probability distribution}}\\
        Uniform & 35811.2 & 36069.0 & 35885.8 & \textbf{35723.5} & \textbf{348.0} & 532.8 & 433.1 & 2588.7 & \textbf{3043.3} & 8398.8 & 6458.2 & 5915.5 \\
        \textit{Runtime} & \textit{24.3} & \textit{8.3} & \textit{5.2} & \textit{0.9} & \textit{8.1} & \textit{1.3} & \textit{1.1} & \textit{0.1} & \textit{7.7} & \textit{0.8} & \textit{1.6} & \textit{0.1} \\
        Beta & 33317.8 & \textbf{32695.8} & 33117.7 & 32746.1 & 169.3 & 112.6 & \textbf{106.8} & 185.2 & 412.6 & \textbf{363.9} & 386.5 & 412.6 \\
        \textit{Runtime} & \textit{27.2} & \textit{7.0} & \textit{5.2} & \textit{0.5} & \textit{2.0} & \textit{0.3} & \textit{1.2} & \textit{0.1} & \textit{0.9} & \textit{0.4} & \textit{1.6} & \textit{0.1} \\
        Normal & 36017.5 & 35779.9 & 36067.6 & \textbf{35616.2} & 352.9 & \textbf{336.2} & 486.4 & 2994.1 & 13403.0 & \textbf{5828.4} & 6555.2 & 5848.0 \\
        \textit{Runtime} & \textit{33.5} & \textit{7.4} & \textit{5.2} & \textit{0.6} & \textit{8.4} & \textit{0.4} & \textit{1.2} & \textit{0.2} & \textit{10.5} & \textit{2.9} & \textit{1.6} & \textit{0.1} \\
        \bottomrule
        \end{tabular}}
    \end{table}

    \begin{table}[htbp]
        \centering
        \caption{EPC and runtime of four algorithms on three network models (node count = 500, without local search)}
        \label{tab:node-500-no-ls}
        \setlength{\tabcolsep}{2.2pt}
        \small
        \resizebox{\textwidth}{!}{%
        \begin{tabular}{l *{3}{*{4}{c}}}       
        \toprule
        & \multicolumn{4}{c}{\textbf{Erdos-Renyi}}
        & \multicolumn{4}{c}{\textbf{Barabasi-Albert}}
        & \multicolumn{4}{c}{\textbf{Watts-Strogatz}} \\
        \cmidrule(lr){2-5}\cmidrule(lr){6-9}\cmidrule(lr){10-13}
        & Greedy & Greedy MIS & Greedy GNN & GNN (1-shot)
        & Greedy & Greedy MIS & Greedy GNN & GNN (1-shot)
        & Greedy & Greedy MIS & Greedy GNN & GNN (1-shot) \\
        \midrule
        \multicolumn{13}{l}{\textbf{Edge probability \(p\)}}\\
        $p = 0.1$ & 60995.7 & \textbf{54924.9} & 57333.8 & 55187.3 & 94.1 & 56.6 & \textbf{45.2} & 60.1 & 111.5 & 105.7 & 95.4 & \textbf{94.4} \\
        \textit{Runtime} & \textit{58.9} & \textit{29.0} & \textit{30.8} & \textit{1.6} & \textit{1.5} & \textit{1.1} & \textit{3.7} & \textit{0.1} & \textit{0.5} & \textit{1.5} & \textit{5.7} & \textit{0.2} \\
        $p = 0.2$ & 96844.0 & 95395.8 & \textbf{94545.8} & 94658.5 & 221.9 & 126.3 & \textbf{108.7} & 171.1 & 366.6 & 294.8 & 281.0 & \textbf{273.7} \\
        \textit{Runtime} & \textit{104.5} & \textit{36.6} & \textit{31.0} & \textit{1.5} & \textit{1.1} & \textit{1.0} & \textit{3.8} & \textit{0.1} & \textit{0.7} & \textit{4.6} & \textit{5.7} & \textit{0.2} \\
        $p = 0.3$ & 100468.8 & 100046.6 & \textbf{99281.7} & 99859.1 & 354.1 & 220.1 & \textbf{203.4} & 442.6 & 1010.8 & 731.1 & 714.1 & \textbf{704.7} \\
        \textit{Runtime} & \textit{114.7} & \textit{35.6} & \textit{32.1} & \textit{1.7} & \textit{5.9} & \textit{2.0} & \textit{3.7} & \textit{0.1} & \textit{4.2} & \textit{3.2} & \textit{5.7} & \textit{0.2} \\
        $p = 0.4$ & 100941.2 & 100685.1 & \textbf{100210.0} & 100763.4 & 474.2 & 371.5 & \textbf{364.1} & 1391.1 & 2582.4 & \textbf{1763.1} & 2352.3 & 2366.0 \\
        \textit{Runtime} & \textit{136.2} & \textit{36.6} & \textit{30.8} & \textit{1.6} & \textit{7.3} & \textit{2.8} & \textit{3.7} & \textit{0.1} & \textit{10.9} & \textit{2.1} & \textit{5.8} & \textit{0.2} \\
        $p = 0.5$ & 100758.3 & \textbf{100396.4} & 100507.7 & 100960.9 & 684.3 & 737.4 & \textbf{670.5} & 5366.9 & 18022.8 & \textbf{8591.9} & 14745.8 & 14378.2 \\
        \textit{Runtime} & \textit{178.0} & \textit{38.3} & \textit{30.8} & \textit{1.6} & \textit{10.6} & \textit{3.1} & \textit{3.7} & \textit{0.2} & \textit{12.8} & \textit{3.8} & \textit{5.7} & \textit{0.2} \\
        $p = 0.6$ & 101009.4 & \textbf{100331.2} & 100670.6 & 101012.3 & 6801.3 & \textbf{1139.8} & 1444.4 & 15947.3 & 66390.9 & \textbf{46467.7} & 54104.6 & 50406.0 \\
        \textit{Runtime} & \textit{259.6} & \textit{27.5} & \textit{30.7} & \textit{1.6} & \textit{12.5} & \textit{2.1} & \textit{3.9} & \textit{0.2} & \textit{23.9} & \textit{5.1} & \textit{5.8} & \textit{0.3} \\
        $p = 0.7$ & 101024.4 & \textbf{100249.0} & 100806.3 & 101024.0 & 14654.5 & \textbf{2975.4} & 3872.8 & 28653.5 & 83467.6 & \textbf{69154.8} & 81705.0 & 76650.4 \\
        \textit{Runtime} & \textit{628.3} & \textit{27.8} & \textit{30.7} & \textit{1.6} & \textit{14.3} & \textit{2.1} & \textit{3.7} & \textit{0.2} & \textit{22.3} & \textit{4.8} & \textit{5.8} & \textit{0.3} \\
        $p = 0.8$ & 101025.0 & \textbf{99666.2} & 100893.7 & 101024.9 & 25683.3 & \textbf{6681.7} & 12279.2 & 40475.9 & 93075.3 & \textbf{84910.2} & 94067.0 & 90534.9 \\
        \textit{Runtime} & \textit{1543.7} & \textit{25.3} & \textit{30.9} & \textit{1.5} & \textit{13.0} & \textit{4.3} & \textit{3.7} & \textit{0.2} & \textit{23.1} & \textit{4.4} & \textit{5.9} & \textit{0.3} \\
        $p = 0.9$ & 101020.6 & 100984.9 & \textbf{100972.7} & 101025.0 & 62166.3 & \textbf{15516.5} & 27307.2 & 51317.1 & \textbf{92887.6} & 94413.8 & 99151.8 & 96628.9 \\
        \textit{Runtime} & \textit{1372.4} & \textit{30.7} & \textit{30.7} & \textit{1.4} & \textit{13.8} & \textit{1.5} & \textit{3.8} & \textit{0.2} & \textit{24.4} & \textit{14.6} & \textit{5.9} & \textit{0.3} \\
        $p = 1.0$ & 101025.0 & \textbf{100111.3} & 101025.0 & 101025.0 & 63636.4 & \textbf{34265.1} & 42085.1 & 60319.9 & 98809.1 & \textbf{98762.2} & 101025.0 & 99184.8 \\
        \textit{Runtime} & \textit{1869.1} & \textit{25.9} & \textit{30.7} & \textit{1.4} & \textit{13.4} & \textit{3.6} & \textit{3.9} & \textit{0.2} & \textit{187.2} & \textit{12.0} & \textit{5.8} & \textit{0.3} \\
        
        \addlinespace
        \multicolumn{13}{l}{\textbf{Edge-probability distribution}}\\
        Uniform & 100930.2 & \textbf{100754.8} & 100908.8 & 100929.2 & 825.2 & 785.3 & \textbf{730.7} & 5720.6 & 22829.5 & \textbf{14012.9} & 20338.8 & 12715.0 \\
        \textit{Runtime} & \textit{137.8} & \textit{34.7} & \textit{30.8} & \textit{1.9} & \textit{10.0} & \textit{1.4} & \textit{3.8} & \textit{0.2} & \textit{11.8} & \textit{2.1} & \textit{5.7} & \textit{0.2} \\
        Beta & 100150.0 & 99919.9 & \textbf{99278.8} & 99528.0 & 244.3 & 262.6 & \textbf{191.0} & 341.3 & 767.6 & 638.4 & 667.1 & \textbf{529.6} \\
        \textit{Runtime} & \textit{91.7} & \textit{39.6} & \textit{30.8} & \textit{1.5} & \textit{3.1} & \textit{0.8} & \textit{3.8} & \textit{0.2} & \textit{1.2} & \textit{1.3} & \textit{5.8} & \textit{0.2} \\
        Normal & 100762.0 & 100699.4 & \textbf{100449.0} & 100944.0 & 908.1 & 746.5 & \textbf{683.3} & 5074.3 & 20302.2 & 11656.5 & 12621.7 & \textbf{10381.0} \\
        \textit{Runtime} & \textit{210.5} & \textit{43.2} & \textit{30.9} & \textit{1.7} & \textit{11.7} & \textit{5.9} & \textit{3.7} & \textit{0.2} & \textit{16.7} & \textit{3.6} & \textit{5.7} & \textit{0.2} \\
        \bottomrule
        \end{tabular}}
    \end{table}

% \begin{acknowledgements}
% If you'd like to thank anyone, place your comments here
% and remove the percent signs.
% \end{acknowledgements}

\noindent\textbf{Data Availability} 
All data generated and examined during this project are accessible at
\url{https://github.com/tuguldur102/SCNDP}.

\section*{Statements and Declarations}

\noindent\textbf{Funding} The authors declare that no funds, grants, or other support were received during the preparation of this manuscript. The authors have no relevant financial or non-financial interests to disclose.

\noindent\textbf{Competing Interests}  The authors have no relevant financial or non-financial interests to disclose.

\noindent\textbf{Author Contributions}  All authors contributed to the study conception and design. All authors read and approved the final manuscript.

\bibliographystyle{plainnat} 
\bibliography{references}   % name your BibTeX data base

\end{document}